\documentclass[12pt]{article}

\usepackage{geometry, fullpage, authblk}
\usepackage{xr-hyper}
\usepackage{times,enumitem}
\usepackage{hyperref}[]
\hypersetup{
	colorlinks=true,
	linkcolor=blue,
	filecolor=magenta,      
	urlcolor=cyan,
	citecolor=blue,
}
\usepackage{ulem}
\usepackage[round]{natbib}
\usepackage{xurl}
\usepackage{bigints}
\usepackage{amsfonts,amsmath,amssymb,amsthm,bbm,bm}
\usepackage{wrapfig}
\usepackage{verbatim,float}
\usepackage{graphicx}
\usepackage{subcaption}
\usepackage[english]{babel}
\usepackage{cancel}
\usepackage{color}
\usepackage[dvipsnames]{xcolor}
\usepackage{cleveref}
\usepackage{multirow}
\usepackage{booktabs}
\usepackage{algorithm}
\usepackage{algorithmic}
\usepackage[flushleft]{threeparttable}
\usepackage{mathtools}

\usepackage{tikz}
\usetikzlibrary{automata, arrows.meta, positioning, fit, shapes}

\title{A Partially Separable Model for Dynamic Valued Networks}

\author[1]{Yik Lun Kei}
\author[2]{Yanzhen Chen}
\author[1]{Oscar Hernan Madrid Padilla}

\affil[1]{\small Department of Statistics, University of California, Los Angeles}
\affil[2]{\small Department of ISOM, Hong Kong University of Science and Technology}

\begin{document}
\maketitle

\begin{abstract}
The Exponential-family Random Graph Model (ERGM) is a powerful model to fit networks with complex structures. However, for dynamic valued networks whose observations are matrices of counts that evolve over time, the development of the ERGM framework is still in its infancy. To facilitate the modeling of dyad value increment and decrement, a Partially Separable Temporal ERGM is proposed for dynamic valued networks. The parameter learning algorithms inherit state-of-the-art estimation techniques to approximate the maximum likelihood, by drawing Markov chain Monte Carlo (MCMC) samples conditioning on the valued network from the previous time step. The ability of the proposed model to interpret network dynamics and forecast temporal trends is demonstrated with real data.
\end{abstract}


\section{Introduction}
\label{sec:intro}

Networks are used to represent relational phenomena in many domains, such as stock relations in financial market \citep{feng2019temporal}, scene graphs in computer vision \citep{suhail2021energy}, and mitochondrial networks in cancer metabolism \citep{han2023spatial}. Conventionally, relations are indicated by the presence or absence of ties. Though connected ties are seemingly identical, relations by nature have degree of strength, which can be represented by generic values to distinguish them. Often, valued networks are dichotomized into binary networks for analysis, which curtails the information that original networks convey. To prevent potential bias from data thresholding \citep{thomas2011valued}, \cite{ValuedERGM} extended the Exponential-family Random Graph Model (ERGM) to fit networks with count dyad values. \cite{desmarais2012statistical} and \cite{wilson2017stochastic} focused on networks with continuous-valued edges. Moreover, \cite{caimo2020multilayer} proposed to model weighted networks in a hierarchical multilayer framework.

Relational phenomena also progress in time. \cite{robins2001random} first proposed to model dynamic networks in a Markovian discrete time framework. \cite{snijders2001statistical} and \cite{snijders2005models} developed a Stochastic Actor-Oriented Model, which is driven by the actor's perspective to make or withdraw ties to other actors. \cite{butts2008} introduced a Relational Event Model, focusing on the action emitted by an entity toward another. Furthermore, \cite{TERGM} defined a Temporal ERGM (TERGM), by specifying a conditional ERGM between consecutive networks. We refer the interested reader to \cite{schaefer2017modeling} for a review in modeling network dynamics.

In this article, we focus on the structure of count valued networks over time. Besides being limited to binary networks, existing frameworks model snapshots of networks, which gives little insight into the underlying dynamic process and little prediction power in how future networks will evolve \citep{jiang2020autoregressive, goyal2020dynamic}. While a snapshot of a valued network presents the structural properties appearing at the observed time point, it does not provide information about the dynamics that produce the structural properties, such as the amount and rate at which the dyad values increase or decrease. Moreover, as we will demonstrate below, without a decomposition that separates dyad value increment and decrement, interpreting network dynamics can be challenging.

\cite{yang2011detecting} proposed a dynamic stochastic block model, by capturing the transition of community memberships for individual nodes. \cite{sewell2016latent} proposed a latent space model, by assuming the probability of a stronger edge between two nodes is greater when they are closer in the latent space. These models provide a good comprehension of the relations between actors over time, though they may not extend to other network structures of interest that signify the generating process. Furthermore, \cite{wyatt2010discovering} included a dynamic feature term in ERGM to capture the change in dyad values over time. Yet the interpretation of network dynamics may be difficult.

The ERGM that defines local forces to shape global structure \citep{hunter2008ergm} is a natural way to model complex networks. Inspired by \cite{STERGM} in modeling dynamic binary networks with a Separable Temporal ERGM (STERGM), we extend the ERGM framework to model dynamic valued networks as follows.

\begin{itemize}

\item We propose a Partially Separable Temporal ERGM (PST ERGM) to fit dynamic valued networks, assuming the factors that increase relational strength are different from those that decrease relational strength. In particular, we construct two intermediate networks to manage dyad value increment and decrement, separately. The dynamics are specified with two sets of network statistics evaluated on the intermediate networks, and we use two sets of parameters to facilitate interpretation.

\item We adapt recent advances in fitting static binary networks from \cite{ImproveERGM} to seed an initial configuration for parameter learning. We exploit the Contrastive Divergence sampling, an abridged MCMC from \cite{hummel2011improving} and \cite{CD}, to expedite the learning process. We also provide the Metropolis-Hastings algorithm to sample dynamic valued networks conditioning on the previous networks.

\item Our experiments show a good performance of the proposed model. In particular, a time-heterogeneous PST ERGM on the students contact networks \citep{mastrandrea2015contact} provides a realistic interpretation of the network dynamics. Furthermore, a time-homogeneous PST ERGM on the baboons interaction networks \citep{gelardi2020measuring} produces reasonable out-of-sample forecasts of the temporal trends.

\end{itemize}

The rest of the paper is organized as follows. In Section \ref{sec:Review}, we review ERGM for static binary and valued networks, and STERGM for dynamic binary networks. In Section \ref{sec:Model_Intro}, we propose the PST ERGM for dynamic valued networks with specifications on the intermediate networks. In Section \ref{sec:Parameter_learning}, we discuss the approximate maximum likelihood estimation and the Metropolis-Hastings algorithm for drawing dynamic valued networks. In Section \ref{sec:Experiments}, we illustrate the methodology with simulated and real data examples. In Section \ref{sec:discussion}, we conclude with a discussion and potential future developments.

\section{ERGM for Networks}
\label{sec:Review}

\subsection{ERGM for Static Binary and Valued Networks}

For a fixed set $N = \{1,2,\cdots,n\}$ of nodes, we can use a network $\bm{y} \in \mathcal{Y}$, in the form of an $n \times n$ matrix, to represent the potential relations for all pairs $(i,j) \in \mathbb{Y} \subseteq N \times N$. The binary networks have dyad $\bm{y}_{ij} \in \{0,1\}$ to represent the absence or presence of a tie and $\mathcal{Y} \subseteq 2^{\mathbb{Y}}$. Let $\mathbb{N}_0$ be the set of natural numbers and $0$. The valued networks have dyad $\bm{y}_{ij} \in \mathbb{N}_0$ to represent the intensity of a tie and $\mathcal{Y} \subseteq \mathbb{N}_0^{\mathbb{Y}}$. We disallow a network to have self-edge, so $\bm{y}_{ij} = 0$ if $i = j$. The relation in a network can be either directed or undirected, where an undirected network has $\bm{y}_{ij} = \bm{y}_{ji}$ for all dyads $(i,j)$. In this article, we focus on undirected networks, and the directed variant follows naturally.

The probabilistic formulation of an ERGM for a network $\bm{y}$ is 
$$
P(\bm{y};\bm{\eta}) = h(\bm{y})\exp[\bm{\eta}^\top \bm{g}(\bm{y}) - \psi(\bm{\eta})],
$$
where $\bm{g}(\bm{y})$, with $\bm{g}: \mathcal{Y} \rightarrow \mathbb{R}^p$, is a vector of network statistics; $\bm{\eta} \in \mathbb{R}^p$ is a vector of unknown parameters; $\exp[\psi(\bm{\eta})] = \sum_{\bm{y}\in\mathcal{Y}} h(\bm{y})\exp[\bm{\eta}^\top \bm{g}(\bm{y})]$ is the normalizing constant; $h(\bm{y})$, with $h: \mathcal{Y} \rightarrow [0,\infty)$, is the reference function. Moreover, the network statistics $\bm{g}(\bm{y})$ may also depend on nodal attributes $\bm{x}$ and dyadic attributes $\bm{z}$. For notational simplicity, we omit the dependence of $\bm{g}(\bm{y})$ on $\bm{x}$ and $\bm{z}$.

In valued ERGM \citep{ValuedERGM}, the parameter space of $\bm{\eta}$ has to ensure that $P(\bm{y};\bm{\eta})$ is a valid probability distribution. When the range of dyad value in a network is $\mathbb{N}_0$, the condition $\exp[\psi(\bm{\eta})] < \infty$ is sufficient to guarantee that $P(\bm{y};\bm{\eta})$ is a valid distribution. Furthermore, the reference function $h(\bm{y})$ underlies a baseline distribution of dyad values. That is when $\bm{\eta} = \bm{0}$, $P(\bm{y};\bm{\eta}) \propto h(\bm{y})$. Specifically, \cite{ValuedERGM} defined a Poisson-reference ERGM and \cite{ergmcount} defined a Binomial-reference ERGM for valued networks with respective reference functions:
$$h_{\text{Pois}}(\bm{y}) = \prod_{(i,j)\in\mathbb{Y}}(\bm{y}_{ij}!)^{-1}
\,\,\,\,\text{and}\,\,\,\,
h_{\text{Bino}}(\bm{y}) = \prod_{(i,j)\in\mathbb{Y}}\binom{m}{\bm{y}_{ij}},$$
where $m$ is a known maximum value that each relationship $\bm{y}_{ij} \in \{0,1,\cdots,m\}$ can take in this network $\bm{y}$. For binary ERGM, usually $h(\bm{y}) = 1$ \citep[e.g.,][]{wasserman1996logit, snijders2002markov, snijders2006new, CERGM}.

\subsection{STERGM for Dynamic Binary Networks}

The TERGM \citep{TERGM} for a binary network $\bm{y}^{t}$ conditional on $\bm{y}^{t-1}$ is
$$P(\bm{y}^{t}|\bm{y}^{t-1};\bm{\eta}) = \exp[\bm{\eta}^\top \bm{g}(\bm{y}^{t},\bm{y}^{t-1}) - \psi(\bm{\eta},\bm{y}^{t-1})],$$
where $\bm{y}^t \in \mathcal{Y}^t \subseteq 2^{\mathbb{Y}}$ is a single network at a discrete time point $t$. The $\bm{g}(\bm{y}^{t},\bm{y}^{t-1})$, with $\bm{g}: \mathcal{Y}^t \times \mathcal{Y}^{t-1} \rightarrow \mathbb{R}^p$, is a vector of network statistics for the transition from $\bm{y}^{t-1}$ to $\bm{y}^t$. Yet \cite{STERGM} demonstrated that higher coefficients in TERGM can lead to inconsistent interpretation of network evolution in terms of incidence and duration. Hence a careful decomposition of network dynamics is needed. In particular, the incidence, how often new ties form, can be measured by dyad formation, and the duration, how long old ties last, can be measured by dyad dissolution.

Instead of modeling the observed $\bm{y}^{t}$ given $\bm{y}^{t-1}$ that muddles network dynamics, \cite{STERGM} designed two intermediate networks, the formation network and dissolution network, between time $t-1$ and $t$ to reflect the incidence and duration. The formation network $\bm{y}^{+,t} \in \mathcal{Y}^{+,t}$ is acquired by adding the edges formed at time $t$ to $\bm{y}^{t-1}$ so $\mathcal{Y}^{+,t} \subseteq \{\bm{y} \in 2^{\mathbb{Y}}: \bm{y} \supseteq \bm{y}^{t-1}\}$. The dissolution network $\bm{y}^{-,t} \in \mathcal{Y}^{-,t}$ is acquired by removing the edges dissolved at time $t$ from $\bm{y}^{t-1}$ so $\mathcal{Y}^{-,t} \subseteq \{\bm{y} \in 2^{\mathbb{Y}}: \bm{y} \subseteq \bm{y}^{t-1}\}$.

Assuming $\bm{y}^{+,t}$ is conditionally independent of $\bm{y}^{-,t}$ given $\bm{y}^{t-1}$, the STERGM \citep{STERGM} for $\bm{y}^t$ conditional on $\bm{y}^{t-1}$ is
\begin{equation}
  P(\bm{y}^t|\bm{y}^{t-1};\bm{\eta}^+,\bm{\eta}^-) = P(\bm{y}^{+,t}|\bm{y}^{t-1};\bm{\eta}^+) \times P(\bm{y}^{-,t}|\bm{y}^{t-1};\bm{\eta}^-),
\label{eq:STERGM}
\end{equation}
with the respective formation model and dissolution model specified as 
\begin{center}
$\begin{aligned}
P(\bm{y}^{+,t}|\bm{y}^{t-1};\bm{\eta}^+) & = \exp[\bm{\eta}^+ \cdot \bm{g}^+(\bm{y}^{+,t},\bm{y}^{t-1}) - \psi(\bm{\eta}^+,\bm{y}^{t-1})]\ \text{and}\\
P(\bm{y}^{-,t}|\bm{y}^{t-1};\bm{\eta}^-) & = \exp[\bm{\eta}^- \cdot \bm{g}^-(\bm{y}^{-,t},\bm{y}^{t-1}) - \psi(\bm{\eta}^-,\bm{y}^{t-1})].
\end{aligned}$
\end{center}
In contrast to the TERGM whose parameter simultaneously influences both incidence and duration, STERGM provides two sets of parameters, where one manages ties formation and the other manages ties dissolution.

The intuition behind the separable parameterization is that the factors and processes that result in ties formation are different from those that result in ties dissolution. Many applications of STERGM on real-world data support the separable assumption for dynamic networks \citep[e.g.,][]{broekel2018disentangling,zhang2019dynamic,xie2020data,uppala2020modeling,Deng2021ABO}. Despite the restriction that the two processes do not interact with each other, substantial improvement in interpretability is gained.

\section{PST ERGM for Dynamic Valued Networks} 
\label{sec:Model_Intro}

\subsection{Increment and Decrement Networks}
\label{subsec:inc_dec}

Although we can use a TERGM \citep{TERGM} to fit dynamic valued networks where $\bm{y}^{t-1}, \bm{y}^t \in \mathbb{N}_0^{\mathbb{Y}}$, parameter interpretation for the network evolution may be difficult. Consider the respective edge sum and stability terms as follows:
$$\bm{g}(\bm{y}^{t}, \bm{y}^{t-1}) = \sum_{(i,j)\in\mathbb{Y}} \bm{y}^t_{ij}
\,\,\,\,\text{and}\,\,\,\,
\bm{g}(\bm{y}^{t}, \bm{y}^{t-1}) = \sum_{(i,j)\in\mathbb{Y}} \mathbbm{1}(\lvert \bm{y}_{ij}^{t} - \bm{y}_{ij}^{t-1} \rvert \leq r),$$
where $r$ is a threshold for the absolute difference in dyad value between time $t-1$ and $t$. In general, a higher coefficient on the edge sum term would favor networks that increase their dyad values at time $t$, while a higher coefficient on the stability term would favor networks that do not change their dyad values at time $t$. In this example, a higher coefficient in the TERGM may lead to inconsistent dyad value movement from time $t-1$ to $t$. Hence, a careful decomposition of the dyad value transition is also needed.

Intuitively, as relational phenomena evolve over time, it is assumed the factors and underlying processes that increase relational strength are different from those that decrease relational strength. For example, in an intensive care unit of a hospital, the number of contacts between a doctor and a patient may increase as the doctor frequently treats the patient during the early stage of infection. The count may further escalate if the patient's symptoms worsen. In contrast, the doctor may reduce the number of contacts as the patient later acquires immunity to the disease. The count may further plummet as the doctor uses medical sensors to monitor the patient after the symptoms alleviate.

Inspired by \cite{STERGM}, we can also design two intermediate networks to consider the dyad value movement, separately. Given two consecutive valued networks $\bm{y}^{t-1}$ and $\bm{y}^t$, we construct an increment network $\bm{y}^{+,t}$ and a decrement network $\bm{y}^{-,t}$ between time $t-1$ and $t$ for a dyad $(i,j)$ with a scaling factor $\beta = 0.5$ as follows:
\begin{equation}
\begin{split}
\label{eq:aug_dim}
   \bm{y}_{ij}^{+,t} & = f^+(\bm{y}_{ij}^{t-1},\bm{y}_{ij}^{t}) \coloneqq 0.5 (\bm{y}_{ij}^{t-1}+\bm{y}_{ij}^{t}) + \beta\lvert\bm{y}_{ij}^{t-1}-\bm{y}_{ij}^{t}\rvert = \max(\bm{y}_{ij}^{t-1},\bm{y}_{ij}^{t})\ \text{and}\\
   \bm{y}_{ij}^{-,t} & = f^-(\bm{y}_{ij}^{t-1},\bm{y}_{ij}^{t}) \coloneqq 0.5 (\bm{y}_{ij}^{t-1}+\bm{y}_{ij}^{t}) - \beta\lvert\bm{y}_{ij}^{t-1}-\bm{y}_{ij}^{t}\rvert = \min(\bm{y}_{ij}^{t-1},\bm{y}_{ij}^{t}).\\
\end{split}
\end{equation}
Note that these are also generalized formulations of the formation network and the dissolution network in \cite{STERGM}, since binary networks are special cases of valued networks in terms of dyad values.

Similar to the formation and dissolution networks, the increment and decrement networks for a dyad $(i,j)$ appear to result in the observed $\bm{y}_{ij}^{t-1}$ and $\bm{y}_{ij}^{t}$. Intrinsically, $\bm{y}_{ij}^{+,t}$ and $\bm{y}_{ij}^{-,t}$ return the values from the average $0.5 (\bm{y}_{ij}^{t-1}+\bm{y}_{ij}^{t})$ tilting away by the absolute difference $\lvert\bm{y}_{ij}^{t-1}-\bm{y}_{ij}^{t}\rvert$ scaled by a factor $\beta$ between two consecutive time points. In this work, we use $\beta=0.5$ to not exaggerate or diminish the absolute difference between the two time points, and to remain on count valued networks as $\bm{y}^{+,t}, \bm{y}^{-,t} \in \mathbb{N}_0^{\mathbb{Y}}$. The resulting $\max$ and $\min$ operations have further implications in model interpretation described in Section \ref{subsec:interpretation}. For $\beta \neq 0.5$ that leads to $\bm{y}^{+,t}, \bm{y}^{-,t} \in \mathbb{R}^{\mathbb{Y}}$, an extension of our framework with the Generalized ERGM \citep{desmarais2012statistical} may be allowed for future development.

\begin{figure}[!htb]
\includegraphics[width=0.55\textwidth]{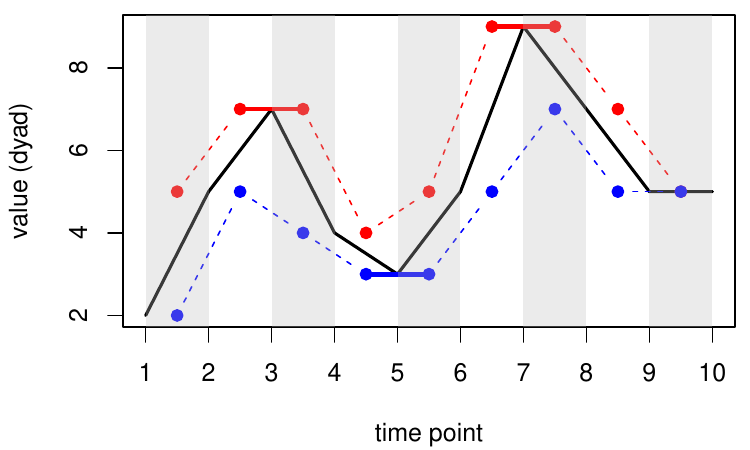}
\centering
\caption{An illustration of $\bm{y}_{ij}^t$ (black) and the constructed $\bm{y}_{ij}^{+,t}$ (red) and $\bm{y}_{ij}^{-,t}$ (blue) over time.}
\label{fig:dyad_value}
\end{figure}

In summary, $\bm{y}^{+,t}$ contains the unchanged dyad values from time $t-1$ to $t$ and those that increased at time $t$, while $\bm{y}^{-,t}$ contains the unchanged dyad values from time $t-1$ to $t$ and those that decreased at time $t$. Furthermore, both of them preserve the momentum when the dyad value starts to change in the opposite direction. As the bolded segments shown in Figure \ref{fig:dyad_value}, the momentum of the changes is delayed to the next interval for a model to digest the stimulus. Similar to \cite{STERGM}, we substitute the sequence of $T$ observed networks with the sequence of $2\times(T-1)$ extracted networks that focus on dyad value movements, as augmented input data to a model. Alternatively, $\bm{y}^{+,t}$ and $\bm{y}^{-,t}$ can be considered as two latent networks that emphasize the transitions between time $t-1$ and $t$, instead of two snapshots of the observed networks $\bm{y}^{t-1}$ and $\bm{y}^{t}$ which give limited information about the dynamics.

\begin{figure}[!htb]
\includegraphics[width=0.45\textwidth]{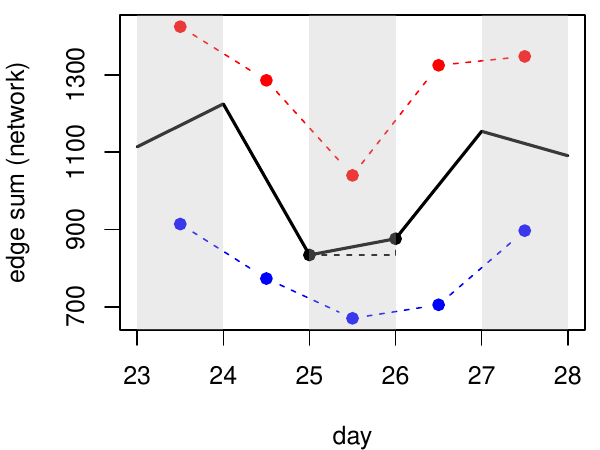}
\centering
\caption{The edge sums of the baboons interaction networks $\bm{y}^t$ (black) from day $23$ to $28$, and the edge sums of the constructed $\bm{y}^{+,t}$ (red) and $\bm{y}^{-,t}$ (blue). The edge sums of $\bm{y}^{25}$ and $\bm{y}^{26}$ are highlighted.}
\label{fig:edge_sum}
\end{figure}

Before proposing the PST ERGM in Section \ref{subsec:model_spec} with details, we further motivate the increment and decrement networks with a comparison between simple fitted models, using the baboons interaction networks \citep{gelardi2020measuring} analyzed in Section \ref{subsec:baboons}. As shown in Figure \ref{fig:edge_sum}, the difference in edge sums $\sum_{(i,j)\in\mathbb{Y}} \bm{y}^t_{ij}$ between $t=25$ and $26$ is relatively small. Fitting a valued ERGM \citep{ValuedERGM} that involves only the edge sum term to $\bm{y}^{26}$, we notice the coefficient is close to that of the same model fitted to $\bm{y}^{25}$. However, fitting the proposed PST ERGM that involves only the edge sum terms to both $\bm{y}^{+,26}$ and $\bm{y}^{-,26}$, we notice the coefficient of $\sum_{(i,j)\in\mathbb{Y}} \bm{y}^{+,26}_{ij}$ is positive and that of $\sum_{(i,j)\in\mathbb{Y}} \bm{y}^{-,26}_{ij}$ is negative. The coefficients of the simple fitted models are displayed in \ref{appendix:comparison}.

In the example with PST ERGM, the positive coefficient indicates an increment among dyad values, while the negative coefficient indicates previously high dyad values tend not to persist to the next time point. There is a fluctuation in dyad values between the observed time points, though the edge sums appear to be unchanged at the observed time points. From $t=25$ to $26$, the total increment of the dyads that increase is $206$, and the total decrement of the dyads that decrease is $164$, resulting in a net increase of $42$ or $11\%$ of total value changes. Without a decomposition that separates dyad value increment and decrement, such dynamics may be neglected. Next, we introduce the proposed PST ERGM in details.

\subsection{Model Specification}
\label{subsec:model_spec}

We first define the form of the model for a sequence of valued networks $\bm{y}^{1}, \cdots, \bm{y}^{T}$ with ERGM specified as the transition between consecutive networks. Under the first order Markov assumption where $\bm{y}^t$ is independent of $\bm{y}^{t-2}, \cdots, \bm{y}^{1}$ conditioning on $\bm{y}^{t-1}$, we have  
\begin{center}
$\begin{aligned}
P(\bm{y}^{T}, \bm{y}^{T-1}, \cdots, \bm{y}^2|\bm{y}^{1}) & = P(\bm{y}^{T}|\bm{y}^{T-1})P(\bm{y}^{T-1}|\bm{y}^{T-2})\cdots P(\bm{y}^{2}|\bm{y}^{1})\\
& = \prod_{t=2}^T h(\bm{y}^{t},\bm{y}^{t-1})\exp[\bm{\eta} \cdot \bm{g}(\bm{y}^{t},\bm{y}^{t-1}) - \psi(\bm{\eta},\bm{y}^{t-1})].
\end{aligned}$
\end{center}
Besides the dynamics between consecutive networks, $P(\bm{y}^1)$ can be specified by a valued ERGM \citep{ValuedERGM} to complete the joint distribution.

To dissect the entanglement between dyad value increment and decrement in dynamic valued networks, it may be straightforward to consider that the increment network $\bm{y}^{+,t}$ is also conditionally independent of the decrement network $\bm{y}^{-,t}$ given $\bm{y}^{t-1}$, as in the STERGM for dynamic binary networks. However, as we will compare the two cases below, a fully separable model for dynamic valued networks can be difficult to obtain, while retaining the information encoded in $\bm{y}^{+,t}$ and $\bm{y}^{-,t}$.

For dynamic binary networks where $\mathcal{Y}^t \subseteq 2^{\mathbb{Y}}$, the STERGM of (\ref{eq:STERGM}) permits us to sample $\bm{y}^{+,t}$ and $\bm{y}^{-,t}$ individually to produce a unique $\bm{y}^t \in \mathcal{Y}^t$. Conditioning on $\bm{y}^{t-1}$ with a particular dyad $\bm{y}_{ij}^{t-1} = 0$, the sampled $\bm{y}_{ij}^{-,t}$ can only be $0$ whereas the sampled $\bm{y}_{ij}^{+,t}$ can be either $0$ or $1$. Once $\bm{y}_{ij}^{+,t}$ is determined, a unique $\bm{y}_{ij}^{t}$ is confirmed. Similarly, conditioning on $\bm{y}^{t-1}$ with a particular dyad $\bm{y}_{ij}^{t-1} = 1$, the sampled $\bm{y}_{ij}^{+,t}$ can only be $1$ whereas the sampled $\bm{y}_{ij}^{-,t}$ can be either $0$ or $1$. Once $\bm{y}_{ij}^{-,t}$ is determined, a unique $\bm{y}_{ij}^{t}$ is confirmed. Therefore, a separable model in the spaces of $\mathcal{Y}^{+,t}$ and $\mathcal{Y}^{-,t}$, proposed by \cite{STERGM}, is a valid probability distribution for a binary network $\bm{y}^t$ conditional on $\bm{y}^{t-1}$.

For dynamic valued networks where $\mathcal{Y}^t \subseteq \mathbb{N}_0^{\mathbb{Y}}$, suppose $P(\bm{y}^t|\bm{y}^{t-1})$ can still be separated into two conditionally independent models as in (\ref{eq:STERGM}), so that we can sample $\bm{y}^{+,t}$ and $\bm{y}^{-,t}$ individually to produce a unique $\bm{y}^t \in \mathcal{Y}^t$. Conditioning on $\bm{y}^{t-1}$ with a particular dyad value $\bm{y}_{ij}^{t-1} \in \mathbb{N}_0$ and under the specification of (\ref{eq:aug_dim}), a sampled $\bm{y}_{ij}^{+,t}$ can be any count value that is greater than or equal to $\bm{y}_{ij}^{t-1}$, and a sampled $\bm{y}_{ij}^{-,t}$ can be any non-negative count value that is smaller than or equal to $\bm{y}_{ij}^{t-1}$. For example, conditioning on $\bm{y}^{t-1}$ with $\bm{y}_{ij}^{t-1} = 3$, if the sampled $\bm{y}_{ij}^{+,t}$ from $P(\bm{y}^{+,t}|\bm{y}^{t-1};\bm{\eta}^+)$ is $5$ and the sampled $\bm{y}_{ij}^{-,t}$ from $P(\bm{y}^{-,t}|\bm{y}^{t-1};\bm{\eta}^-)$ is $2$, a unique $\bm{y}_{ij}^{t}$ is unidentifiable given the two intermediate dyad values. The separated generating processes cannot decide whether the dyad value $\bm{y}_{ij}^{t-1} = 3$ should increase to $\bm{y}_{ij}^{t} = 5$ or decrease to $\bm{y}_{ij}^{t} = 2$ at time $t$.

Since the TERGM in our framework can no longer be separated into two conditionally independent models as in \cite{STERGM}, our proposed Partially Separable Temporal ERGM (PST ERGM) for a sequence of valued networks is
\begin{equation}
\label{eq:PSTERGM}
\begin{split}
\prod_{t=2}^T P(\bm{y}^t|\bm{y}^{t-1};\bm{\eta}) & = \prod_{t=2}^T h(\bm{y}^{t},\bm{y}^{t-1})\exp[\bm{\eta} \cdot \bm{g}(\bm{y}^{t},\bm{y}^{t-1}) - \psi(\bm{\eta},\bm{y}^{t-1})]\\
& = \prod_{t=2}^T h^+(\bm{y}^{+,t}) h^-(\bm{y}^{-,t}) \frac{\exp[\bm{\eta}^+ \cdot \bm{g}^+(\bm{y}^{+,t},\bm{y}^{t-1}) + \bm{\eta}^- \cdot \bm{g}^-(\bm{y}^{-,t},\bm{y}^{t-1})]}{\exp[\psi(\bm{\eta}^+,\bm{\eta}^-,\bm{y}^{t-1})]},
\end{split}
\end{equation}
with $h(\bm{y}^{t},\bm{y}^{t-1}) = h^+(\bm{y}^{+,t}) \times h^-(\bm{y}^{-,t}) \in \mathbb{R}$ and $\bm{\eta} = (\bm{\eta}^+,\bm{\eta}^-) \in \mathbb{R}^p$. The network statistics $\bm{g}(\bm{y}^{t},\bm{y}^{t-1}) = \big(\bm{g}^+(\bm{y}^{+,t},\bm{y}^{t-1}),\bm{g}^-(\bm{y}^{-,t},\bm{y}^{t-1})\big) \in \mathbb{R}^p$ is a concatenation of the increment network statistics $\bm{g}^+(\bm{y}^{+,t},\bm{y}^{t-1}) \in \mathbb{R}^{p_1}$ and the decrement network statistics $\bm{g}^-(\bm{y}^{-,t},\bm{y}^{t-1}) \in \mathbb{R}^{p_2}$ such that $p_1 + p_2 = p$. The normalizing constant $\exp[\psi(\bm{\eta}^+,\bm{\eta}^-,\bm{y}^{t-1})]$ is
$$\sum_{\bm{y}^t \in\mathcal{Y}^t} h^+(\bm{y}^{+,t}) \exp[\bm{\eta}^+ \cdot \bm{g}^+(\bm{y}^{+,t},\bm{y}^{t-1})] \times h^-(\bm{y}^{-,t}) \exp[\bm{\eta}^- \cdot  \bm{g}^-(\bm{y}^{-,t},\bm{y}^{t-1})].$$
Though we cannot generate $\bm{y}^{+,t}$ and $\bm{y}^{-,t}$ to produce a unique $\bm{y}^t$ with PST ERGM, we can directly sample $\bm{y}^t$ by using the Metropolis-Hastings algorithm described in Section \ref{subsec:MH}.

In this article, we use the Poisson and Binomial reference functions:
\begin{equation}
\label{eq:reference_measure}
h^+(\bm{y}^{+,t}) = \prod_{(i,j)\in\mathbb{Y}}(\bm{y}_{ij}^{+,t}!)^{-1}
\,\,\,\,\text{and}\,\,\,\, 
h^-(\bm{y}^{-,t}) = \prod_{(i,j)\in\mathbb{Y}}\binom{m^t}{\bm{y}_{ij}^{-,t}},
\end{equation}
for the increment and decrement process, respectively. The term $m^t$ in $h^-(\bm{y}^{-,t})$ is a pre-determined maximum value that each dyad value $\bm{y}^{-,t}_{ij} \in \{0,1,\cdots,m^t\}$ can take in $\bm{y}^{-,t}$. Furthermore, the reference function $h^+(\bm{y}^{+,t})$ in the increment process does not require an upper bound for a dyad value $\bm{y}_{ij}^{+,t}$ that it can increase to, but $\bm{y}_{ij}^{+,t}$ has an implicit lower bound that is equal to $\bm{y}_{ij}^{t-1}$ inherited from the construction of $\bm{y}_{ij}^{+,t} = \max(\bm{y}_{ij}^{t-1},\bm{y}_{ij}^{t})$. In the decrement process, the reference function $h^-(\bm{y}^{-,t})$ imposes an upper bound $m^t$ for each dyad value $\bm{y}_{ij}^{-,t}$ that it can decrease from, with an explicit lower bound that is equal to $0$.

To capture the variation in structural properties between different intervals, we can also specify a time-heterogeneous PST ERGM $\prod_{t=2}^T P(\bm{y}^t|\bm{y}^{t-1};\bm{\eta}^{t})$ as
\begin{center}
$\begin{aligned}
\prod_{t=2}^T \frac{h^+(\bm{y}^{+,t}) \exp[\bm{\eta}^{+,t} \cdot \bm{g}^+(\bm{y}^{+,t},\bm{y}^{t-1})] \times h^-(\bm{y}^{-,t}) \exp[\bm{\eta}^{-,t} \cdot  \bm{g}^-(\bm{y}^{-,t},\bm{y}^{t-1})]}{\exp[\psi(\bm{\eta}^{+,t},\bm{\eta}^{-,t},\bm{y}^{t-1})]},
\end{aligned}$
\end{center}
where $\bm{\eta}^t = (\bm{\eta}^{+,t},\bm{\eta}^{-,t})$ differs by time $t$. Unless otherwise noted, we focus on the time-homogeneous PST ERGM of (\ref{eq:PSTERGM}) whose parameter $\bm{\eta} = (\bm{\eta}^{+},\bm{\eta}^{-})$ is fixed across $t = 2, \cdots, T$. The time-heterogeneous PST ERGM is a special case of (\ref{eq:PSTERGM}) as $\bm{\eta}^t = (\bm{\eta}^{+,t},\bm{\eta}^{-,t})$ can be learned sequentially for each $t$.

\subsubsection{Reference Measures}

Inherited from valued ERGM \citep{ValuedERGM}, the reference function $h(\bm{y}^{t},\bm{y}^{t-1})$ specified with (\ref{eq:reference_measure}) in a PST ERGM underlies a baseline distribution for $\bm{y}^t$. Consider a PST ERGM with the edge sums of increment and decrement networks as two network statistics:
$$\bm{g}(\bm{y}^t, \bm{y}^{t-1}) = \big(\bm{g}^+(\bm{y}^{+,t},\bm{y}^{t-1}),\bm{g}^-(\bm{y}^{-,t},\bm{y}^{t-1})\big) = \Big(\sum_{(i,j)\in\mathbb{Y}} \bm{y}_{ij}^{+,t}, \sum_{(i,j)\in\mathbb{Y}} \bm{y}_{ij}^{-,t}\Big) \in \mathbb{R}^2.$$
Let $\mathcal{Y}^t(\bm{y}^{t-1}) \subseteq \{\bm{y}^t \in \mathbb{N}_0^{\mathbb{Y}}: \bm{y}_{ij}^{t} > \bm{y}_{ij}^{t-1}\ \forall\ (i,j) \in \mathbb{Y}\}$ be a sample space for $\bm{y}^t$ starting from $\bm{y}^{t-1}$. The increment network $\bm{y}^{+,t}$ is essentially the $\bm{y}^t$, and the PST ERGM $P(\bm{y}^{t}|\bm{y}^{t-1};\bm{\eta})$ for $\bm{y}^t \in \mathcal{Y}^t(\bm{y}^{t-1})$ becomes a dyadic independent truncated Poisson distribution:
$$\prod_{(i,j)\in\mathbb{Y}} \frac{ (\bm{y}_{ij}^{t}!)^{-1} \exp(\bm{\eta}^{+} \cdot \bm{y}_{ij}^{t})}{\exp\big(\exp(\bm{\eta}^{+})\big) - \sum_{u=0}^{\bm{y}_{ij}^{t-1}} (u!)^{-1} \exp(\bm{\eta}^{+} \cdot u)} = \prod_{(i,j)\in\mathbb{Y}} \frac{P_{\text{Pois}}\big(\bm{y}_{ij}^t\big)}{1 - \sum_{u=0}^{\bm{y}_{ij}^{t-1}} P_{\text{Pois}}(u)},$$
where $P_{\text{Pois}}(x)$ denotes the probability mass function of ${\text{Poisson}}\big(\lambda = \exp(\bm{\eta}^+)\big)$ evaluated at $x$. Moreover, let $\mathcal{Y}^t(\bm{y}^{t-1}) \subseteq \{\bm{y}^t \in \mathbb{N}_0^{\mathbb{Y}}: \bm{y}_{ij}^{t} < \bm{y}_{ij}^{t-1} \leq m^t \ \forall\ (i,j) \in \mathbb{Y}\}$ be another sample space for $\bm{y}^t$. The decrement network $\bm{y}^{-,t}$ is essentially the $\bm{y}^t$, and the PST ERGM $P(\bm{y}^{t}|\bm{y}^{t-1};\bm{\eta})$ for $\bm{y}^t \in \mathcal{Y}^t(\bm{y}^{t-1})$ becomes a dyadic independent truncated Binomial distribution:
$$\prod_{(i,j)\in\mathbb{Y}} \frac{ \binom{m^t}{\bm{y}^{t}_{ij}} \exp(\bm{\eta}^{-} \cdot \bm{y}_{ij}^{t})}{ \big(1+\exp(\bm{\eta}^-)\big)^{m^t} -  \sum_{u=\bm{y}_{ij}^{t-1}}^{m^t} \binom{m^t}{u}\exp(\bm{\eta}^{-} \cdot u)} = \prod_{(i,j)\in\mathbb{Y}} \frac{P_{\text{Bino}}\big(\bm{y}_{ij}^t\big)}{1 - \sum_{u=\bm{y}_{ij}^{t-1}}^{m^t} P_{\text{Bino}}\big(u\big)},$$
where $P_{\text{Bino}}(x)$ denotes the probability mass function of ${\text{Binomial}}\big(m^t, p = \text{logit}^{-1}(\bm{\eta}^-)\big)$ evaluated at $x$. The derivations of the two special cases are provided in \ref{appendix:special_cases}.

Next, we consider the general network statistics $\bm{g}(\bm{y}^{t},\bm{y}^{t-1}) \in \mathbb{R}^p$. For $\mathcal{Y}^t(\bm{y}^{t-1}) \subseteq \{\bm{y}^t \in \mathbb{N}_0^{\mathbb{Y}}: \bm{y}_{ij}^{t} > \bm{y}_{ij}^{t-1}\ \forall\ (i,j) \in \mathbb{Y}\}$, the PST ERGM becomes a Poisson-reference valued ERGM:
$$P(\bm{y}^{t}|\bm{y}^{t-1};\bm{\eta}) = \frac{\big( \prod_{(i,j)\in\mathbb{Y}}(\bm{y}_{ij}^{t}!)^{-1} \big) \exp[\bm{\eta}^+ \cdot \bm{g}^+(\bm{y}^{t},\bm{y}^{t-1})]}{\sum_{\bm{y}^t \in \mathcal{Y}^t(\bm{y}^{t-1})} \big( \prod_{(i,j)\in\mathbb{Y}}(\bm{y}_{ij}^{t}!)^{-1} \big) \exp[\bm{\eta}^+ \cdot \bm{g}^+(\bm{y}^{t},\bm{y}^{t-1})]},$$
with the Poisson reference function and the increment network statistics directly evaluated at $\bm{y}^t \in \mathcal{Y}^t(\bm{y}^{t-1})$. Moreover, for $\mathcal{Y}^t(\bm{y}^{t-1}) \subseteq \{\bm{y}^t \in \mathbb{N}_0^{\mathbb{Y}}: \bm{y}_{ij}^{t} < \bm{y}_{ij}^{t-1} \leq m^t \ \forall\ (i,j) \in \mathbb{Y}\}$, the PST ERGM becomes a Binomial-reference valued ERGM:
$$P(\bm{y}^{t}|\bm{y}^{t-1};\bm{\eta}) = \frac{\big( \prod_{(i,j)\in\mathbb{Y}}\binom{m^t}{\bm{y}^{t}_{ij}} \big) \exp[\bm{\eta}^- \cdot \bm{g}^-(\bm{y}^{t},\bm{y}^{t-1})]}{\sum_{\bm{y}^t \in \mathcal{Y}^t(\bm{y}^{t-1})} \big( \prod_{(i,j)\in\mathbb{Y}}\binom{m^t}{\bm{y}^{t}_{ij}} \big) \exp[\bm{\eta}^- \cdot \bm{g}^-(\bm{y}^{t},\bm{y}^{t-1})]},$$
with the Binomial reference function and the decrement network statistics directly evaluated at $\bm{y}^t \in \mathcal{Y}^t(\bm{y}^{t-1})$. Next, we provide four levels of interpretation to PST ERGM.

\subsection{Model Interpretation}
\label{subsec:interpretation}

We first compare the formulations of STERGM and the proposed PST ERGM. The STERGM \citep{STERGM} for dynamic binary networks given as 
$$P(\bm{y}^{t}|\bm{y}^{t-1};\bm{\eta})  = \frac{\exp[\bm{\eta}^+ \cdot \bm{g}^+(\bm{y}^{+,t},\bm{y}^{t-1})]}{\exp[\psi(\bm{\eta}^+,\bm{y}^{t-1})]} \times \frac{\exp[\bm{\eta}^- \cdot \bm{g}^-(\bm{y}^{-,t},\bm{y}^{t-1})]}{\exp[\psi(\bm{\eta}^-,\bm{y}^{t-1})]}$$
is fully separable, while the PST ERGM for dynamic valued networks given as
$$P(\bm{y}^t|\bm{y}^{t-1};\bm{\eta}) \propto\ \exp[\bm{\eta}^+ \cdot \bm{g}^+(\bm{y}^{+,t},\bm{y}^{t-1})] \times \exp[\bm{\eta}^- \cdot  \bm{g}^-(\bm{y}^{-,t},\bm{y}^{t-1})]$$ 
is partially separable. Though both models are distributions over the observed networks, the user-specified network statistics $\bm{g}^+$ and $\bm{g}^-$ in both models are evaluated at $\bm{y}^{+,t}$ and $\bm{y}^{-,t}$ that are extracted from the observed networks. Hierarchically, two layers of network features are extracted: dyad value movements via $\bm{y}^{+,t}$ and $\bm{y}^{-,t}$, and their structural properties via $\bm{g}^+$ and $\bm{g}^-$. These dynamics are then captured by $\bm{g}(\bm{y}^t, \bm{y}^{t-1}) = \big(\bm{g}^+(\bm{y}^{+,t},\bm{y}^{t-1}),\bm{g}^-(\bm{y}^{-,t},\bm{y}^{t-1})\big)$ with an exponential-family model. In alignment with the separability, the choice of network statistics in $\bm{g}^+(\bm{y}^{+,t},\bm{y}^{t-1}) \in \mathbb{R}^{p_1}$ can be different from that in $\bm{g}^-(\bm{y}^{-,t},\bm{y}^{t-1}) \in \mathbb{R}^{p_2}$, depending on the user's knowledge of which local forces matter in which process to shape the global structures over time. In contrast to STERGM, the increment and decrement processes in the PST ERGM are not conditionally independent, as discussed in Section \ref{subsec:model_spec}.

Next, we present the similarity between STERGM and PST ERGM in how they dissect network evolution. The idea of separability originates from epidemiology to approximate disease dynamics: Prevalence $\approx$ Incidence $\times$ Duration. \cite{STERGM} used the formation and dissolution networks to reflect incidence (how often new ties are formed) and duration (how long old ties last since they were formed). Since the duration of ties is the inverse of the rate at which ties dissolve, the parameter $\bm{\eta}^-$ in the dissolution model of STERGM can signify the persistence of network features \citep{STERGM}. Translating these into PST ERGM, the structural properties of dynamic valued networks are characterizations of the amount and rate of dyad value increment and decrement. The more often dyad values increase and they increase with a greater magnitude per time step, the more high dyad values will be presented over time. The less frequent dyad values decrease and they decrease with a smaller magnitude per time step, the more high dyad values will be preserved over time. Conceptually, we can broadly regard incidence as how often dyad values increase, and regard duration as how long dyad values have kept increasing until they decrease. The duration of the continuing increment is the inverse of the rate at which dyad values decrease. Inherited from Equation (\ref{eq:aug_dim}), the increment and decrement networks are also encoded with the values to which the dyads have increased or decreased.

Furthermore, the ERGM framework has a dyadic level interpretation for a static binary network in terms of change statistics \citep{hunter2008ergm}. We provide similar interpretation for dynamic valued networks with PST ERGM, by changing a dyad value in $\bm{y}^t$ to see how increment and decrement processes impact the network structures. Specifically, we calculate the ratio of probabilities of two networks that are identical except for a single dyad. Suppose the dyad value $\bm{y}_{ij}^t \in \mathbb{N}_0$ jumps from $a$ to $b$ where $b \neq a$. Conditioning on the rest of the network $\bm{y}_{-ij}^t$ and $\bm{y}^{t-1}$, the ratio $P(\bm{y}_{ij}^t = b | \bm{y}_{-ij}^t , \bm{y}^{t-1}) / P(\bm{y}_{ij}^t = a | \bm{y}_{-ij}^t , \bm{y}^{t-1})$ is
$$\frac{\bm{y}^{+,t}_{\text{old}}!}{\bm{y}^{+,t}_{\text{new}}!} \exp[\bm{\eta}^{+} \cdot \Delta \bm{g}^+(\bm{y}^{+,t},\bm{y}^{t-1})_{ij}] \times \binom{m^t}{\bm{y}^{-,t}_{\text{new}}} / \binom{m^t}{\bm{y}^{-,t}_{\text{old}}}  \exp[\bm{\eta}^{-} \cdot \Delta \bm{g}^-(\bm{y}^{-,t},\bm{y}^{t-1})_{ij}].$$
The change statistics $\Delta \bm{g}^+(\bm{y}^{+,t},\bm{y}^{t-1})_{ij}$ denote the difference between $\bm{g}^+(\bm{y}^{+,t},\bm{y}^{t-1})$ with $\bm{y}^{+,t}_{ij}= \bm{y}_{\text{new}}^{+,t}$ and $\bm{g}^+(\bm{y}^{+,t},\bm{y}^{t-1})$ with $\bm{y}^{+,t}_{ij}=\bm{y}_{\text{old}}^{+,t}$ while rest of the $\bm{y}^{+,t}$ remains the same, and the $\Delta \bm{g}^-(\bm{y}^{-,t},\bm{y}^{t-1})_{ij}$ are denoted similarly except for notational difference. When both $a,b > \bm{y}_{ij}^{t-1}$, only $\bm{y}^{+,t}$ by construction is updated, regardless of $a > b$ or $b > a$. In other words, only the increment process contributes to the structural changes in $\bm{y}^t$ when the dyad value $\bm{y}_{ij}^t$ is different. Similarly, when both $a,b < \bm{y}_{ij}^{t-1}$, only the decrement process contributes to the structural changes in $\bm{y}^t$. However, when the value $b$ falls on the other side of $\bm{y}_{ij}^{t-1}$ with respect to the value $a$, both increment and decrement processes start to contribute to the structural changes in $\bm{y}^t$. Intuitively, the construction by (\ref{eq:aug_dim}) can be considered as rectified linear units, gated by $\bm{y}_{ij}^{t-1}$ that differs by dyad $(i,j)$ and time point $t$. When a dyad value overcomes a threshold, it activates the corresponding process via user-specified network statistics to impact the network structures.

In addition to the dyadic level interpretation, the parameters of PST ERGM can be interpreted at the structural level, as in \cite{STERGM}. Though we cannot generate $\bm{y}^{+,t}$ and $\bm{y}^{-,t}$ to produce $\bm{y}^t$ given fixed parameters, we can learn the unknown parameters of PST ERGM given observed $\bm{y}^{t}$ and $\bm{y}^{t-1}$ to interpret the dynamics via $\bm{g}(\bm{y}^t, \bm{y}^{t-1})$ from an exponential-family model. When fitting a PST ERGM on the observed $\bm{y}^t$ conditional on $\bm{y}^{t-1}$, the dyad value movements (increased, decreased, or unchanged) between consecutive time points become fixed. As we augment the observed networks with a sequence of $2 \times (T-1)$ networks that separate dyad value increment and decrement, the learned parameters $(\bm{\eta}^+, \bm{\eta}^-) \in \mathbb{R}^p$ can signify the structural changes in $\bm{y}^t$ stemming from the two processes. In general, for a particular positive $\bm{g}^+_i(\bm{y}^{+,t},\bm{y}^{t-1})$ in the increment process, a positive $\bm{\eta}^+_i$ is associated with increasing dyad values to have more instances of the feature that is tracked by $\bm{g}^+_i(\bm{y}^{+,t},\bm{y}^{t-1})$ in the extracted $\bm{y}^{+,t}$. On the contrary, a negative $\bm{\eta}^+_i$ will disrupt the emergence of this feature by not increasing dyad values, resulting in fewer instances of the feature in $\bm{y}^{+,t}$. For a particular positive $\bm{g}^-_i(\bm{y}^{-,t},\bm{y}^{t-1})$ in the decrement process, a positive $\bm{\eta}^-_i$ is associated with not decreasing dyad values to have more instances of the feature that is tracked by $\bm{g}^-_i(\bm{y}^{-,t},\bm{y}^{t-1})$ in the extracted $\bm{y}^{-,t}$. However, a negative $\bm{\eta}^-_i$ will target this feature by reducing dyad values, resulting in fewer instances of the feature in $\bm{y}^{-,t}$. Equivalently, a negative $\bm{\eta}^-_i$ is associated with a shorter duration of the feature appearance. Moreover, since we learn the parameters $(\bm{\eta}^+, \bm{\eta}^-)$ jointly as described in Section \ref{sec:Parameter_learning}, the parameters that reflect the dynamics via $\bm{g}(\bm{y}^t, \bm{y}^{t-1}) = \big(\bm{g}^+(\bm{y}^{+,t},\bm{y}^{t-1}),\bm{g}^-(\bm{y}^{-,t},\bm{y}^{t-1})\big)$ balance the two processes. Though we assume the factors that increase relational strength are different from those that decrease relational strength, they can be interacting in practice and the effects of interactions over time are absorbed into $(\bm{\eta}^+,\bm{\eta}^-)$ with a partially separable model.

\begin{figure}[!htb]
\centering

\begin{tikzpicture} [node distance = 2.8 cm, on grid, state/.style={circle, draw, minimum size=1.4 cm}]

\node (y1) [state] {$\bm{y}^{1}$};
\node (y2+) [state, dashed, thick, above right = of y1] {$\bm{y}^{+,2}$};
\node (y2-) [state, dotted, thick, below right = of y1] {$\bm{y}^{-,2}$};
\node (y2) [state, below right = of y2+] {$\bm{y}^{2}$};

\node (yT-1) [state, right = of y2] {$\bm{y}^{T-1}$};
\node (yT+) [state, dashed, thick, above right = of yT-1] {$\bm{y}^{+,T}$};
\node (yT-) [state, dotted, thick, below right = of yT-1] {$\bm{y}^{-,T}$};
\node (yT) [state, below right = of yT+] {$\bm{y}^{T}$};

\node (eta+) [state, regular polygon, regular polygon sides=3, dashed, thick, above right = of yT, fill=lightgray] {$\tilde{\bm{\eta}}^{+}$};
\node (eta-) [state, regular polygon, regular polygon sides=3, dotted, thick, below right = of yT, fill=lightgray, shape border rotate=180] {$\tilde{\bm{\eta}}^{-}$};
\node (yT+1) [state, below right = of eta+, fill=lightgray] {$\hat{\bm{y}}^{T+1}$};

\node[draw, fit={(y2+) (yT+)}, rounded corners, label=center:Increment Process, label=above:$\prod_{t=2}^T h^+(\bm{y}^{+,t}) \exp\big(\bm{\eta}^+ \cdot \bm{g}^+(\bm{y}^{+,t}{,}\ \bm{y}^{t-1})\big)$] (y2T+) {};

\node[draw, fit=(y2-) (yT-), rounded corners, label=center:Decrement Process, label=below:$\prod_{t=2}^T h^-(\bm{y}^{-,t}) \exp\big(\bm{\eta}^- \cdot \bm{g}^-(\bm{y}^{-,t}{,}\ \bm{y}^{t-1})\big)$] (y2T-) {};

\node[draw, fit=(yT) (eta+) (eta-) (yT+1), rounded corners, label=above:Forecasting Process, ] (prediction) {};

\path [-stealth, thick]
    (y1) edge node {}  (y2+)
    (y1) edge node {}  (y2-)
    (y2) edge node {}  (y2+)
    (y2) edge node {}  (y2-)
    
    (y2) -- node[auto=false]{\ldots} (yT-1)
    
    (yT-1) edge node {}  (yT+)
    (yT-1) edge node {}  (yT-)
    (yT) edge node {}  (yT+)
    (yT) edge node {}  (yT-)
    
    (yT) edge node [above] {MCMC} node[below] {Sampling}(yT+1)
    (y2T+) edge [dashed] node[above] {Jointly} node[below] {Learning} (eta+)
    (y2T-) edge [dotted] node[above] {Jointly} node[below] {Learning} (eta-)
    (eta+) edge [dashed, bend left] node {} (yT+1)
    (eta-) edge [dotted, bend right] node {} (yT+1)
    ;
    
\end{tikzpicture}

\caption{An overview of PST ERGM for dynamic valued networks.}
\label{fig:STERGM_Overview}
\end{figure}
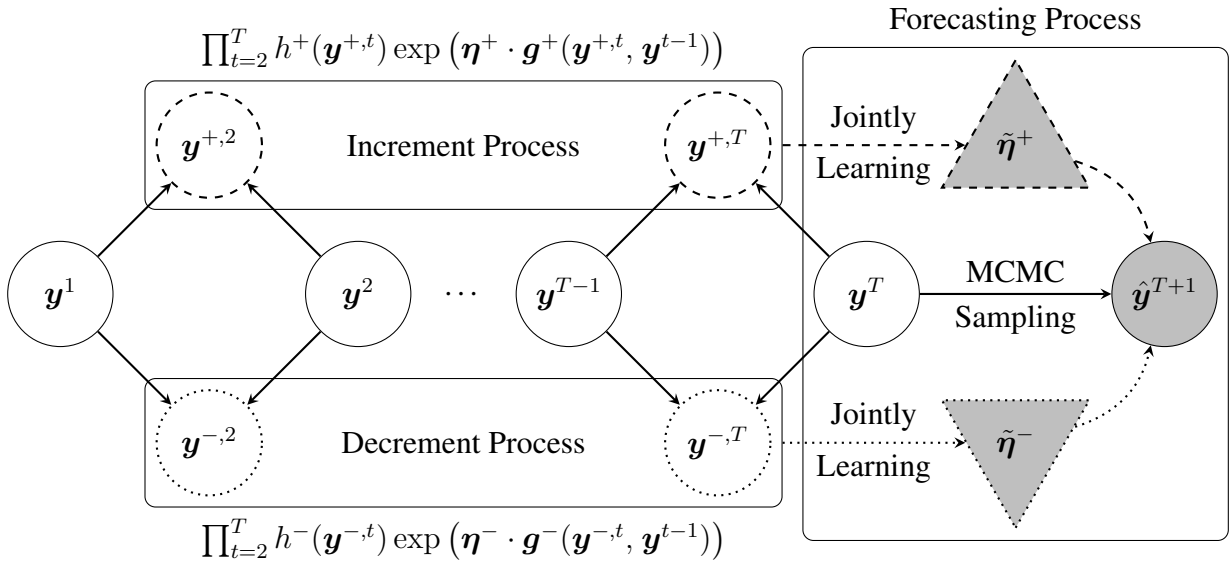

Figure \ref{fig:STERGM_Overview} gives an overview of the PST ERGM framework. The white solid circles denote the sequence of observed networks as time passes from left to right. The dashed circles denote the sequence of increment networks, and the dotted circles denote the sequence of decrement networks. Note that each observed network is utilized multiple times to extract information that emphasizes the transition between consecutive time steps. The model with respect to the observed networks is partially separated into the increment process and the decrement process. Once the parameters in the two triangles are learned jointly, we can perform MCMC sampling to generate $\hat{\bm{y}}^{T+1}$ in the forecasting process. Though $\bm{y}^t$ can be further conditioned on more previous networks to calculate the network statistics and to construct $\bm{y}^{+,t}$ and $\bm{y}^{-,t}$, we only discuss PST ERGM under first order Markov assumption in this article.

\section{Likelihood-Based Inference} 
\label{sec:Parameter_learning}

The PST ERGM parameter estimation consists of two phases, extended from recent advances in fitting static binary networks. Specifically, we first maximize the log-likelihood ratio to seed an initial configuration, followed by the Newton-Raphson method to refine the parameters. The algorithms are provided in \ref{appendix:algo}.

\subsection{Log-likelihood Ratio}

Throughout, the parameters $(\bm{\eta}^+,\bm{\eta}^-) = \bm{\eta} \in \mathbb{R}^p$ are estimated jointly. The log-likelihood of PST ERGM in (\ref{eq:PSTERGM}) with $\big(\bm{g}^+(\bm{y}^{+,t},\bm{y}^{t-1}),\bm{g}^-(\bm{y}^{-,t},\bm{y}^{t-1})\big) = \bm{g}(\bm{y}^{t},\bm{y}^{t-1}) \in \mathbb{R}^p$ is
$$l(\bm{\eta}) = \sum_{t=2}^T \Big\{ \log[h(\bm{y}^{t},\bm{y}^{t-1})] + \bm{\eta} \cdot \bm{g}(\bm{y}^{t},\bm{y}^{t-1}) - \psi(\bm{\eta},\bm{y}^{t-1})\Big\}.$$
The term $\exp[\psi(\bm{\eta},\bm{y}^{t-1})]$ involves a sum over all possible networks in $\mathcal{Y}^{t} \subseteq \mathbb{N}_0^{\mathbb{Y}}$, which is often computationally intractable except for models with particular conditional independence properties \citep{lauritzen2018random} or small networks \citep{yon2021exponential}. Consequently, we approximate the MLE using MCMC methods. To maximize the log-likelihood, we calculate its first and second derivative with respect to $\bm{\eta}$:
\begin{equation}
\label{eq:Hessian}
\bm{S}(\bm{\eta}) = \sum_{t=2}^T \Big\{\bm{g}(\bm{y}^{t},\bm{y}^{t-1}) - \mathbb{E}_{\bm{\eta}}[\bm{g}(\bm{y}^{t},\bm{y}^{t-1})]\Big\}
\,\,\,\,\text{and}\,\,\,\, 
\bm{H}(\bm{\eta}) = \sum_{t=2}^T -\text{Cov}[\bm{g}(\bm{y}^{t},\bm{y}^{t-1})].
\end{equation}
The gradient $\bm{S}(\bm{\eta})$ illustrates that ERGM fitting is essentially a feature pursuit: finding a parameter $\bm{\eta}$ such that the expected network statistics are close to the observed network statistics. Moreover, to obtain the standard errors of $\bm{\eta}$, the Fisher Information matrix can be approximated by the Hessian as $\bm{I}(\bm{\eta}) \approx - \hat{\bm{H}}(\tilde{\bm{\eta}})$ evaluated at the learned parameter $\tilde{\bm{\eta}}$ with MCMC samples \citep{CERGM}.

Approximating $\mathbb{E}_{\bm{\eta}}[\bm{g}(\bm{y}^{t},\bm{y}^{t-1})]$ and $\text{Cov}[\bm{g}(\bm{y}^{t},\bm{y}^{t-1})]$ of (\ref{eq:Hessian}) with MCMC samples, the parameter $\bm{\eta}$ can be updated iteratively by the Newton-Raphson method. However, generating new MCMC samples at each learning iteration is computationally expensive. To reduce the computational burden, we use the log-likelihood ratio as a new objective function to approximate the MLE as in \cite{snijders2002markov}, and \cite{CERGM}. Let $\bm{\eta}_0$ be another initialized parameter, the log-likelihood ratio is
$$r(\bm{\eta},\bm{\eta}_0) = \sum_{t=2}^T \Big\{ (\bm{\eta}-\bm{\eta}_0)^\top \bm{g}(\bm{y}^{t},\bm{y}^{t-1})- \log \mathbb{E}_{\bm{\eta}_0} \big(\exp[(\bm{\eta} - \bm{\eta}_0)^\top \bm{g}(\bm{y}^{t},\bm{y}^{t-1})]\big)\Big\}.$$
Note that the distribution to draw samples from is now changed as we introduce an initialized parameter $\bm{\eta}_0$ to the log-likelihood ratio. 

Generating a sufficiently large number of samples from $P(\bm{y}^{t}|\bm{y}^{t-1};\bm{\eta}_0)$ only once for each time $t$, and then iterating a Newton-Raphson method with respect to $\bm{\eta}$ until convergence yields a maximizer of the approximated log-likelihood ratio. Though anchored on pre-determined samples can greatly expedite the estimation, the efficiency of not having to update the MCMC samples between learning iterations comes with a cost. \cite{geyer1992constrained} pointed out that the approximated log-likelihood ratio via MCMC samples degrades quickly as $\bm{\eta}_0$ moves away from $\bm{\eta}$. We address this issue in the next section.

\subsection{Normality Approximation and Partial Stepping}

\cite{ImproveERGM} proposed two amendments to improve the fitting for static binary networks. We adapt them to the PST ERGM for dynamic valued networks, to seed a starting point for the Newton-Raphson method. Let $\bm{y}^{t}_{1}, \cdots, \bm{y}^{t}_{s}$ be a list of $s$ networks sampled from $P(\bm{y}^{t}|\bm{y}^{t-1};\bm{\eta}_0)$. Since they are drawn from the same distribution, we can assume their network statistics multiplied by the difference of the two parameters, $(\bm{\eta}-\bm{\eta}_0)^\top \bm{g}(\bm{y}^{t},\bm{y}^{t-1})$, follow a normal distribution $\mathcal{N}(\mu_t, \sigma_t^2)$ with
$$\mu_t = (\bm{\eta}-\bm{\eta}_0)^\top \bm{\mu}_{t}
\,\,\,\,\text{and}\,\,\,\,
\sigma_t^2 = (\bm{\eta}-\bm{\eta}_0)^\top \bm{\Sigma}_{t} (\bm{\eta}-\bm{\eta}_0).$$
The $\bm{\mu}_{t}$ and $\bm{\Sigma}_{t}$ are the respective mean vector and covariance matrix of $\bm{g}(\bm{y}^{t},\bm{y}^{t-1})$ evaluated from the sampled $\bm{y}^{t}_{1}, \cdots, \bm{y}^{t}_{s}$. Given that $\exp[(\bm{\eta} - \bm{\eta}_0)^\top \bm{g}(\bm{y}^{t},\bm{y}^{t-1})]$ is now log-normally distributed, the ratio of the two normalizing constants at $t$ can be replaced by
$$\mathbb{E}_{\bm{\eta}_0} \big(\exp[(\bm{\eta} - \bm{\eta}_0)^\top \bm{g}(\bm{y}^{t},\bm{y}^{t-1})]\big) \approx \exp(\mu_t + \frac{1}{2}\sigma_t^2),$$
and the approximated log-likelihood ratio becomes
$$\hat{r}_s(\bm{\eta},\bm{\eta}_0) = (\bm{\eta}-\bm{\eta}_0)^\top \Big[\sum_{t=2}^T \bm{g}(\bm{y}^{t},\bm{y}^{t-1}) - \sum_{t=2}^T \bm{\mu}_{t}\Big] - \frac{1}{2}(\bm{\eta}-\bm{\eta}_0)^\top \Big[\sum_{t=2}^T \bm{\Sigma}_{t}\Big] (\bm{\eta}-\bm{\eta}_0).$$

Although the approximated log-likelihood ratio $\hat{r}_s(\bm{\eta},\bm{\eta}_0)$ degrades quickly as $\bm{\eta}_0$ moves away from $\bm{\eta}$, we can restrict the amount of parameter update to prevent the degradation of $\hat{r}_s(\bm{\eta},\bm{\eta}_0)$. With a step length $\gamma^t \in (0,1]$ at time $t$, we create a pseudo-observation
\begin{equation}
\label{eq:pseudo-observation}
\hat{\bm{\xi}}(\bm{y}) = \sum_{t=2}^T \gamma^t \bm{g}(\bm{y}^{t},\bm{y}^{t-1}) + \sum_{t=2}^T (1 - \gamma^t) \bm{\mu}_{t}
\end{equation}
in between the observed network statistics $\sum_{t=2}^T \bm{g}(\bm{y}^{t},\bm{y}^{t-1})$ and the estimated network statistics $\sum_{t=2}^T \bm{\mu}_{t}$ from MCMC samples drawn from $P(\bm{y}^{t}|\bm{y}^{t-1};\bm{\eta}_0)$. Instead of the difference between $\sum_{t=2}^T \bm{\mu}_{t}$ and $\sum_{t=2}^T \bm{g}(\bm{y}^{t},\bm{y}^{t-1})$, we limit the amount of parameter update based on the difference between $\sum_{t=2}^T \bm{\mu}_{t}$ and $\hat{\bm{\xi}}(\bm{y})$ in each learning iteration. Empirically, the step length $\gamma^t$ is helpful in dampening the possibly drastic variations across different time intervals, for a stable parameter update when searching for an initial configuration.

Thus, we sequentially update the parameter in the direction of the MLE, while maintaining the approximated log-likelihood ratio $\hat{r}_s(\bm{\eta},\bm{\eta}_0)$ estimated by MCMC samples is reasonably accurate. The closed-form solution for the maximizer of the approximated log-likelihood ratio at a specific learning iteration is
$$\tilde{\bm{\eta}} = \bm{\eta}_0 + \Big[\sum_{t=2}^T \bm{\Sigma}_{t}\Big]^{-1} \Big[\hat{\bm{\xi}}(\bm{y}) - \sum_{t=2}^T \bm{\mu}_{t}\Big].$$
Once an initial configuration is obtained from maximizing the approximated log-likelihood ratio, we proceed to the Newton-Raphson method to further update the parameter near its convergence. In this two-phase fusion where each component performs its designated task, both procedures require shorter learning iterations and undertake smaller computational burdens than only on their own.

\subsection{MCMC for Dynamic Valued Networks}
\label{subsec:MH}

Fitting PST ERGM can be heavily dependent on MCMC sampling. In this section, we introduce the Metropolis-Hastings algorithm for drawing $\bm{y}^{t}$ conditional on $\bm{y}^{t-1}$. The superscript $t$ is omitted for $\bm{y}^t$, $\bm{y}^{+,t}$, $\bm{y}^{-,t}$, and $m^t$ to facilitate notational simplicity, as MCMC sampling is performed within a particular time point $t$. Additionally, we use a superscript $k$ to refer to the current MCMC iteration.

In practice, valued networks are often sparse. To accommodate the sparsity of networks, as our proposal distribution, we employ a zero-inflated Poisson distribution that is also used in \texttt{ergm.count} \citep{ergmcount}, an \texttt{R} library for static valued networks:
\begin{equation}
  P(\bm{y}^{k+1}_{ij}; \lambda, \pi_0) =
    \begin{cases}
      \pi_0 + (1 - \pi_0) \exp(-\lambda) & \text{if $\bm{y}^{k+1}_{ij}=0$};\\
      (1-\pi_0) \exp(-\lambda) \times \lambda^{\bm{y}^{k+1}_{ij}}/\bm{y}^{k+1}_{ij}! & \text{if $\bm{y}^{k+1}_{ij} \in \mathbb{N}$},\\
    \end{cases}      
\label{eq:proposed}
\end{equation}
where $\lambda = \bm{y}^{k}_{ij}+0.5$ and $\pi_0 \in [0,1)$ is a pre-defined probability for the proposed dyad jumping to 0. The $0.5$ in $\lambda$ prevents the proposed $\bm{y}^{k+1}_{ij}$ from locking into $0$ when $\bm{y}^{k}_{ij} = 0$, and the proposed $(i,j)$ is chosen randomly. We let $\pi_0 = 0.2$, a default value for the Poisson proposal distribution used in \texttt{ergm.count}, and it can be adjusted based on the user's prior knowledge on the sparsity of networks. The acceptance ratio $\alpha$ for the proposed $\bm{y}_{ij}^{k+1}$ is
\begin{equation}
q \times \frac{\bm{y}^{+,k}_{ij}!}{\bm{y}^{+,k+1}_{ij}!} \exp[{\bm{\eta}^+} \cdot \Delta \bm{g}^+(\bm{y}^{+},\bm{y}^{t-1})_{ij}] \times \binom{m}{\bm{y}^{-,k+1}_{ij}} / \binom{m}{\bm{y}^{-,k}_{ij}}  \exp[{\bm{\eta}^-} \cdot \Delta \bm{g}^-(\bm{y}^{-},\bm{y}^{t-1})_{ij}],
\label{eq:acceptance}
\end{equation}
where $\bm{y}^{+}$ and $\bm{y}^{-}$ are constructed from the observed $\bm{y}^{t-1}$ and the proposed network at MCMC iteration $k+1$. The change statistics $\Delta \bm{g}^+(\bm{y}^+,\bm{y}^{t-1})_{ij}$ denote the difference between $\bm{g}^+(\bm{y}^+,\bm{y}^{t-1})$ with $\bm{y}^+_{ij}= \bm{y}^{+,k+1}_{ij}$ and $\bm{g}^+(\bm{y}^+,\bm{y}^{t-1})$ with $\bm{y}^+_{ij}=\bm{y}^{+,k}_{ij}$ while rest of the $\bm{y}^+$ remains the same. The change statistics $\Delta \bm{g}^-(\bm{y}^-,\bm{y}^{t-1})_{ij}$ are calculated similarly except for notational difference, and the transition probability ratio $q$ is
$$q = P(\bm{y}^{k}_{ij}; \lambda = \bm{y}^{k+1}_{ij}+0.5,\pi_0) / P(\bm{y}^{k+1}_{ij}; \lambda =  \bm{y}^{k}_{ij}+0.5,\pi_0).$$

In this context, we propose a dyad value from the space of $\bm{y}^t$, but we decide to accept the proposed dyad value based on the construction of increment network $\bm{y}^{+,t}$ and decrement network $\bm{y}^{-,t}$, namely the dynamics between time $t-1$ and $t$. As we consolidate the temporal aspect into the PST ERGM, MCMC sampling becomes especially important in forecasting future networks besides its primary usage in parameter learning. Conditioning on the last observed network $\bm{y}^{T}$ under first order Markov assumption, we can forecast $\hat{\bm{y}}^{T+1}$ given the learned parameters $\tilde{\bm{\eta}}^+$ and $\tilde{\bm{\eta}}^-$ with the above scheme.

\subsubsection{Contrastive Divergence Sampling}

\cite{hummel2011improving} applied a $K$-step Contrastive Divergence ($\text{CD}_K$) sampling, an abridged MCMC, to speed up parameter estimation for static binary networks. Introduced in \cite{hinton2002training} and \cite{carreira2005contrastive}, and applied to ERGM in \cite{fellows2014} and \cite{CD}, the Contrastive Divergence (CD) for ERGM is formulated as
$$\text{CD}_K = \text{KL}[P_{\text{data}}(\bm{y}^{\text{obs}})\ \Vert\ P_{\infty}(\bm{y})] - \text{KL}[P_{K}(\bm{y})\ \Vert\ P_{\infty}(\bm{y})].$$
The $P_{\text{data}}(\bm{y}^{\text{obs}})$ is the distribution of the observed data, $P_{\infty}(\bm{y})$ is the true model distribution, and $P_{K}(\bm{y})$ is the distribution of $K$-step MCMC samples \citep{hummel2011improving}. The gradient of $\text{CD}_K$ for minimization given as 
$$\nabla \text{CD}_K = \bm{g}(\bm{y}^{\text{obs}}) - \mathbb{E}_{K}[\bm{g}(\bm{y})] = \bm{0}$$
builds the foundation of $\text{CD}_K$ sampling, where $\mathbb{E}_{K}[\bm{g}(\bm{y})]$ is the expected network statistics under the distribution of $K$-step MCMC samples.

In $\text{CD}_K$ sampling, each sampled network is generated after $K$ transitions starting from the observed network, so a burn-in phase is not required, and a tremendous sample size is not indispensable. Moreover, a small value of $K$ can be used. In seeding an initial configuration via maximizing the approximated log-likelihood ratio, the $\text{CD}_K$ sampling is in favor of the normality approximation, since each sampled network is at most $K$ dyads different from the observed network. In the second phase where the learned parameter is close to the MLE, the network statistics of the MCMC samples are close to those of the observed networks. However, a small value of $K$ generates a pseudo-observation of (\ref{eq:pseudo-observation}) that is not distinct from $\sum_{t=2}^T \bm{g}(\bm{y}^{t},\bm{y}^{t-1})$, which may compromise the advantage of the partial stepping. Hence, a trade-off among the number of transitions, sample size, and learning iteration is needed. See \cite{CD} for a detailed study of using CD in ERGM fitting, especially regarding the choice of hyperparameters and stopping criterion.

\section{Experiments} 
\label{sec:Experiments}

In this section, we apply PST ERGM to simulated and real data, for demonstrative purpose. Though the following real data can be analyzed by other frameworks such as Stochastic Actor-Oriented Model \citep{snijders2001statistical} or Relational Event Model \citep{butts2008}, we focus on the structure of dynamic valued networks, instead of the instantaneous action emitted by an actor given time-ordered sequences of historical events. In practice, when a network with relational strengths between nodes is observed at multiple time points, we can apply PST ERGM to investigate the significance of network structures over time, especially those that can signify the network generating process.

The network statistics of interest are chosen from an extensive list in \texttt{ergm} \citep{ergm_package}, an \texttt{R} library for network analysis. In this demonstration, the choice of network statistics in the increment process is identical to that in the decrement process. For real data experiments, the detail of the implementation and the formulations of selected network statistics are provided in \ref{appendix:exp}.

\subsection{Simulation Study}

In this simulation with $T=10$ and $n=50$, we test our Metropolis-Hastings algorithm by generating $\bm{y}^{2},\dots,\bm{y}^T$ given $\bm{y}^1$ with pre-defined parameters $\bm{\eta}^+$ and $\bm{\eta}^-$. We then test our parameter learning algorithms by estimating the coefficients from the artificial data to compare with the true parameters. We choose four network statistics, (1) edge sum, (2) zeros, (3) mutuality, (4) transitive weight for both increment and decrement processes. The $\bm{g}^+(\bm{y}^{+,t},\bm{y}^{t-1}) \in \mathbb{R}^4$ is specified as follows:
$$\Big( \sum_{ij} \bm{y}_{ij}^{+,t},\ \sum_{ij} \mathbbm{1}(\bm{y}_{ij}^{+,t} = 0),\ \sum_{i<j} \sqrt{\bm{y}_{ij}^{+,t}\bm{y}_{ji}^{+,t}},\ \sum_{ij} \min (\bm{y}_{ij}^{+,t}, \max_{k \in N}(\min(\bm{y}_{ik}^{+,t},\bm{y}_{kj}^{+,t}))) \Big).$$
The $\bm{g}^-(\bm{y}^{-,t},\bm{y}^{t-1}) \in \mathbb{R}^4$ is specified similarly except for notational differences.

\begin{figure}[!htb]
\includegraphics[width=1\textwidth]{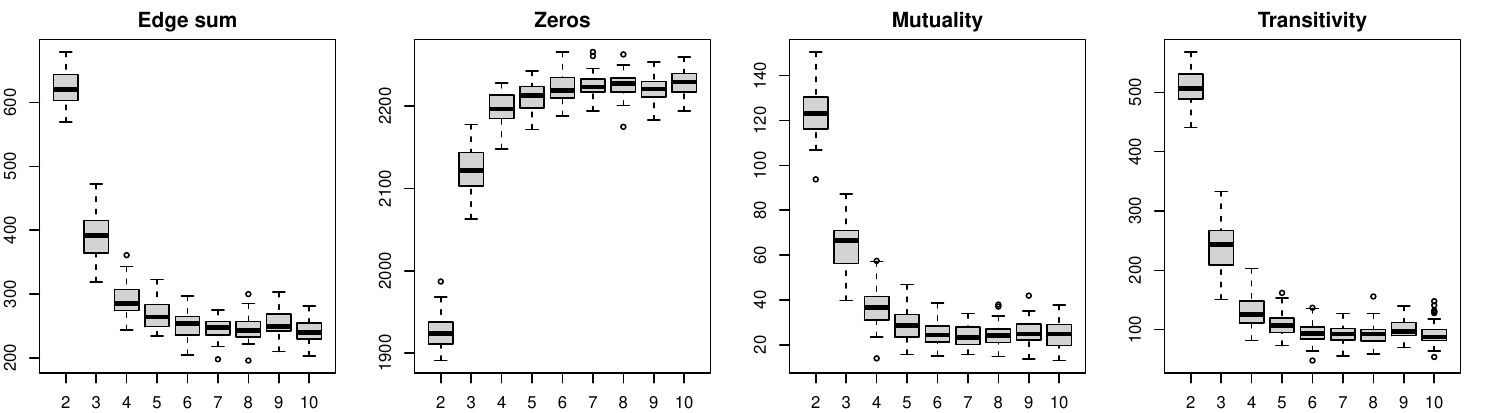}
\centering
\caption{The distributions of network statistics based on $50$ generated sequences of networks. The network statistics are evaluated at $\bm{y}^t$ from $t=2$ to $10$.}
\label{fig:boxplot_sparse}
\end{figure}

We initialize $\bm{\eta}^+ = (-2, 2, 1, 1)$, $\bm{\eta}^- = (-1, 2, 1, 1)$, and the maximum dyad value for decrement networks $m^t=3$ for $t = 2,\dots,T$. To ensure the networks are sampled with reasonable mixing and do not depend on initialization, each sampled $\bm{y}^{t}$ is generated after $20 \times n \times n$ MCMC transitions starting from an empty network. We repeat the process until we have $50$ sequences of $\bm{y}^1, \cdots, \bm{y}^T$. As shown in Figure \ref{fig:boxplot_sparse}, the simulated network statistics have converged both within and across time points. In particular, the simulated networks are designed to be sparse, about $80\%$ of the dyad values in $\bm{y}^t$ are zeros.

We then learn the parameter of PST ERGM for each generated sequence. To seed an initial configuration, we apply $20$ iterations of partial stepping starting from a zero vector. An MCMC sample size of $100$ with $\text{CD}_n$ sampling is used for each time $t$. The number of MCMC transitions is set to $n$. Subsequently, to refine the parameter, we apply $5$ iterations of Newton-Raphson method, where an MCMC sample size of $1000$ with $\text{CD}_n$ sampling is used for each time $t$. The medians and standard deviations of $|\tilde{\bm{\eta}}^+ - \bm{\eta}^+|$ over $50$ estimations are reported in Table~\ref{tbl:result_inc}. The corresponding results for $|\tilde{\bm{\eta}}^- - \bm{\eta}^-|$ are reported in Table~\ref{tbl:result_dec}.

\begin{table}[!htbp]
    \centering
    \begin{tabular}{ l  c  c  c  c }
\hline
\# of time step \& node & $|\tilde{\bm{\eta}}_1^+ - \bm{\eta}_1^+|$ & $|\tilde{\bm{\eta}}_2^+ - \bm{\eta}_2^+|$ & $|\tilde{\bm{\eta}}_3^+ - \bm{\eta}_3^+|$  & $|\tilde{\bm{\eta}}_4^+ - \bm{\eta}_4^+|$\\
\hline
$T=10, n=50$ & 0.0027 (0.065) & 0.0046 (0.087) & 0.0021 (0.056) & 0.0128 (0.064)\\
\hline
\end{tabular}
    \caption{
    The medians (standard deviation) of  $|\tilde{\bm{\eta}}^+ - \bm{\eta}^+|$ over $50$ estimations.}
    \label{tbl:result_inc}
\end{table}

\begin{table}[!htbp]
    \centering
    \begin{tabular}{ l  c  c  c  c }
\hline
\# of time step \& node & $|\tilde{\bm{\eta}}_1^- - \bm{\eta}_1^-|$ & $|\tilde{\bm{\eta}}_2^- - \bm{\eta}_2^-|$ & $|\tilde{\bm{\eta}}_3^- - \bm{\eta}_3^-|$ & $|\tilde{\bm{\eta}}_4^- - \bm{\eta}_4^-|$\\
\hline
$T=10, n=50$ & 0.0471 (0.169) & 0.0228 (0.182) & 0.0108 (0.162) & 0.0033 (0.076)\\
\hline
\end{tabular}
    \caption{
    The medians (standard deviation) of  $|\tilde{\bm{\eta}}^- - \bm{\eta}^-|$ over $50$ estimations.}
    \label{tbl:result_dec}
\end{table}

\begin{figure}[!htb]
\includegraphics[width=1\textwidth]{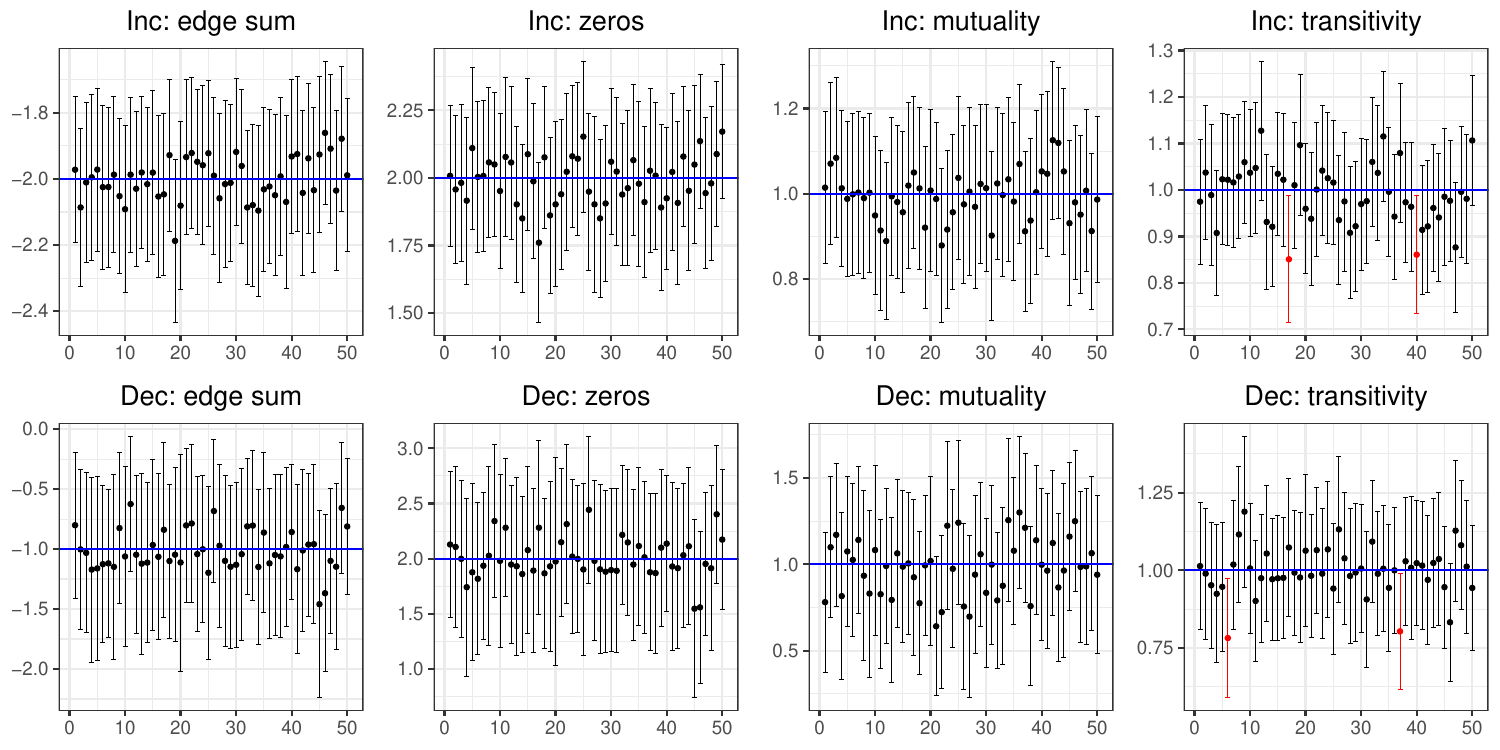}
\centering
\caption{The $95\%$ confidence intervals (black bars) of the $50$ learned parameters (dots) for each network statistic. The blue horizontal lines indicate the true parameter values $\bm{\eta}^+ = (-2, 2, 1, 1)$ and $\bm{\eta}^- = (-1, 2, 1, 1)$. The red bars indicate the confidence intervals that do not cover the true parameters.}
\label{fig:sim_sparse}
\end{figure}

On average, the estimations are close to the true parameters as the medians of absolute differences are close to zeros. We also check if the $95\%$ confidence intervals of the learned parameters cover the true parameters. Figure \ref{fig:sim_sparse} displays the confidence intervals $\tilde{\bm{\eta}}_i \pm 1.96 \tilde{\bm{\sigma}}_{i}$ for the $50$ estimations, where $\tilde{\bm{\sigma}}_{i}$ denotes the standard error of $\tilde{\bm{\eta}}_i$. The standard errors are obtained from the Fisher Information matrix $\bm{I}(\bm{\eta}) \approx - \hat{\bm{H}}(\tilde{\bm{\eta}})$ of (\ref{eq:Hessian}) evaluated at the learned parameter $\tilde{\bm{\eta}}$ with $100$ sampled networks for each $t$. Each sampled network is generated after $1000$ MCMC transitions starting from the observed $\bm{y}^t$. We notice that the true parameters are covered by the confidence intervals most of the time.

\subsection{Modeling: Students Contact Networks}
\label{subsec:friendship}

\cite{mastrandrea2015contact} used wearable sensors to detect face-to-face contacts between students among nine classes in a high school. The real-time contact events were logged for every $20$-second interval of any two students within a physical distance of $1.5$ meters from 02-Dec-2013 to 06-Dec-2013. Additionally, online social network (Facebook) was submitted by the students voluntarily. In this demonstration, we model student interactions within one of the nine classes, whose class name is MP. There are $n=29$ students which consist of $11$ females and $18$ males. We divide the entries by day to construct $T=5$ undirected valued networks, where $\bm{y}^t_{ij}$ is the number of unique contacts between student $i$ and student $j$ on day $t$. The duration of each contact can be different and expansive. The nodal covariate $\bm{x}_i \in \{\text{F}, \text{M}\}$ is the gender of student $i$, and dyadic covariate $\bm{z}_{ij} \in \{1,0\}$ indicates whether student $i$ and student $j$ are friends on Facebook or not.

We choose six network statistics of interest for analysis, and we learn a time-heterogeneous PST ERGM $\prod_{t=2}^5 P(\bm{y}^t|\bm{y}^{t-1};\bm{\eta}^t)$ for the data. A time-homogeneous model was attempted, but the large variation between different intervals suggests that a time-heterogeneous model is appropriate and realistic. The estimated coefficients and standard errors for the increment process are reported in Table~\ref{tbl:Friendship_inc}. The corresponding results for the decrement process are reported in Table~\ref{tbl:Friendship_dec}.

\begin{table}[!htb]
\centering
\begin{threeparttable}
    
\begin{tabular}{ l  r  r  r  r }
\hline
\multicolumn{1}{l}{Network Statistics} &  
\multicolumn{1}{c}{$\bm{\eta}^{+,2}$} & \multicolumn{1}{c}{$\bm{\eta}^{+,3}$} &
\multicolumn{1}{c}{$\bm{\eta}^{+,4}$} & \multicolumn{1}{c}{$\bm{\eta}^{+,5}$} \\
\hline
Edge sum          & \textbf{3.215} (0.092)  & \textbf{3.316} (0.097)  & \textbf{3.023} (0.111)  & \textbf{3.197} (0.079) \\
Dispersion        & \textbf{-6.825} (0.231) & \textbf{-7.576} (0.201) & \textbf{-6.470} (0.222) & \textbf{-6.506} (0.201)\\
Homophily (M)     & \textbf{0.151} (0.073)  & \textbf{0.223} (0.079)  & 0.096 (0.086)           & \textbf{0.135} (0.059)\\
Heterophily (M-F) &  0.042 (0.068)          &  0.083 (0.080)          & 0.067 (0.082)           & 0.056 (0.057)\\
Facebook          & -0.005 (0.042)          & \textbf{0.123} (0.031)  & -0.033 (0.047)          & \textbf{0.092} (0.035)\\
Transitive weight & \textbf{-0.142} (0.038) & \textbf{-0.068} (0.026) & -0.015 (0.044)          & \textbf{-0.187} (0.032)\\
\hline
\end{tabular}

\begin{tablenotes}
\small
\item Coefficients statistically significant at $0.05$ level are bolded.
\end{tablenotes}

\end{threeparttable}
\caption{The parameter estimation (standard error) of $\bm{\eta}^{+,t}$ for the students contact networks.}
\label{tbl:Friendship_inc}
\end{table}

\begin{table}[!htb]
\centering
\begin{threeparttable}
    
\begin{tabular}{ l  r  r  r  r }
\hline
\multicolumn{1}{l}{Network Statistics} &  
\multicolumn{1}{c}{$\bm{\eta}^{-,2}$} & \multicolumn{1}{c}{$\bm{\eta}^{-,3}$} &
\multicolumn{1}{c}{$\bm{\eta}^{-,4}$} & \multicolumn{1}{c}{$\bm{\eta}^{-,5}$}\\
\hline
Edge sum          & \textbf{0.432} (0.214)  & \textbf{-0.980} (0.220) & \textbf{-0.630} (0.183) & -0.294 (0.162)\\
Dispersion        & \textbf{-6.572} (0.277) & \textbf{-6.144} (0.323) & \textbf{-6.102} (0.275) & \textbf{-6.680} (0.270)\\
Homophily (M)     & -0.161 (0.184)          & \textbf{1.344} (0.213)  & \textbf{0.665} (0.189)  & \textbf{0.390} (0.150)\\
Heterophily (M-F) & -0.222 (0.190)          & \textbf{0.481} (0.174)  &  -0.205 (0.189)         & 0.100 (0.152)\\
Facebook          & -0.083 (0.088)          & \textbf{0.231} (0.105)  & \textbf{0.193} (0.088)  & 0.099 (0.079)\\
Transitive weight & 0.087 (0.066)           & \textbf{-0.445} (0.075) & \textbf{-0.446} (0.044) & -0.077 (0.045)\\
\hline
\end{tabular}

\begin{tablenotes}
\small
\item Coefficients statistically significant at $0.05$ level are bolded.
\end{tablenotes}

\end{threeparttable}
\caption{The parameter estimation (standard error) of $\bm{\eta}^{-,t}$ for the students contact networks.}
\label{tbl:Friendship_dec}
\end{table}

The positive coefficients on the edge sum term in the increment process from $t=2$ to $5$ indicate frequent interactions among students throughout the week. However, the negative coefficients on the edge sum term in the decrement process from $t=3$ to $5$ suggest a short duration on the increase of contact occurrences. In other words, the number of contacts fluctuates over time, which supports the adoption of a time-heterogeneous model. Similarly, the highly negative coefficients on the dispersion term in both increment and decrement processes suggest a strong degree of over-dispersion in the number of contacts. This can be verified by the standard deviation of non-zero dyad values across five days, which is $10.2$, whereas the mean is about half in magnitude, which is $5.8$.

In the increment process from $t=2$ to $5$, the positive coefficients on the homophily for males suggest a strong gender effect in promoting more interactions among male students. Additionally, the positive coefficients in the decrement process from $t=3$ to $5$ indicate that the active interactions among males tend to be ongoing once they have begun. However, these effects are less significant for students of different genders, given the imbalanced proportion between females and males. As supporting evidence, about $89\%$ or $136$ out of $\binom{18}{2} = 153$ pairs of male students have contact events logged by sensors, and their average span is $2.8$ days. In contrast, about $71\%$ or $141$ out of $198$ pairs of male and female students have contact events logged, and their average span is $2.3$ days.

Furthermore, in the increment process from $t=2$ to $5$, the alternating signs of coefficients on the Facebook term indicate that if two students are friends online, they occasionally have active interactions in school. However, the majority of positive coefficients on the Facebook term in the decrement process suggest that online friendships maintain offline interactions. About $82\%$ or $119$ out of $145$ reported Facebook friendships have contact events logged in school, and their average span is $2.9$ days. Lastly, the transitive relationship in the number of contacts is weak, as indicated by the negative coefficients in the increment process. The majority of negative coefficients in the decrement process suggests that the transitivity tends not to persist over time.

\begin{figure}[!htbp]
\includegraphics[width=\textwidth]{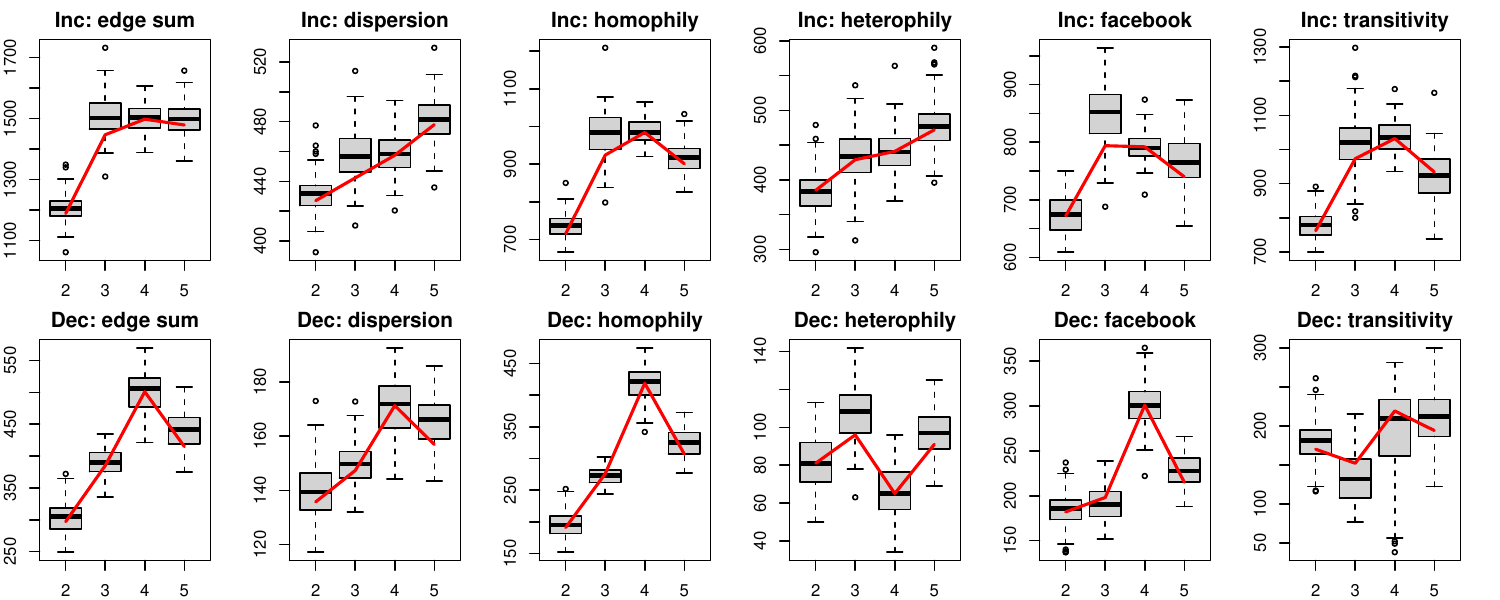}
\centering
\caption{The distribution of the sampled network statistics (box plots) and the observed network statistics values (red lines) across four consecutive intervals for increment (Inc) and decrement (Dec) processes.}
\label{fig:friendship}
\end{figure}

To validate the learned model heuristically, we simulate networks with the estimated parameters to compare the sampled network statistics with the observed network statistics. For $t=2, \dots, 5$, we generate $100$ valued networks $\bm{y}^t$ conditional on the observed $\bm{y}^{t-1}$, where each sampled network is generated after $200 \times n \times n$ MCMC transitions starting from an empty network. We then construct the corresponding $\bm{y}^{+,t}$ and $\bm{y}^{-,t}$ and calculate their network statistics. The distributions of the simulated network statistics and the observed network statistics values are displayed in Figure \ref{fig:friendship}. Overall, the simulated network statistics align with the observed values, suggesting the learned PST ERGM is a good representation of the observed data in terms of the six selected network statistics. 

\subsection{Forecasting: Baboons Interaction Networks}
\label{subsec:baboons}

\cite{gelardi2020measuring} studied the interactions among $n = 13$ baboons for a duration of $28$ days. The contact events are recorded by sensors for every 20-second interval of any two primates within proximity of $1.5$ meters. The data is divided by day to construct a sequence of $T = 28$ undirected valued networks, where $\bm{y}_{ij}^t$ is the number of unique contacts between baboon $i$ and baboon $j$ on day $t$. The duration of each contact can be different and expansive. In this experiment, we learn a time-homogeneous PST ERGM based on the data from day $1$ to $23$, and we forecast $5$ subsequent networks to compare with the observed network from day $24$ to $28$. Though a time-heterogeneous model can learn the transitions well, it lacks the ability to forecast future networks as a learned parameter $\bm{\eta}^t$ is tailored to the designated time point $t$ and cannot be extended to the next time point $t+1$.

\begin{figure}[!htbp]
\includegraphics[width=0.7\textwidth]{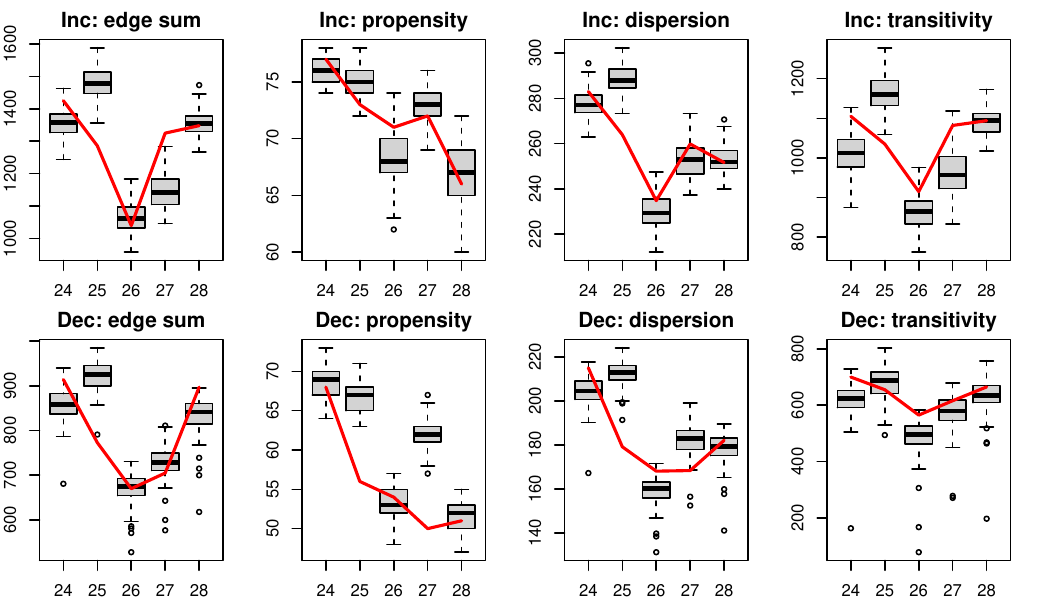}
\centering
\caption{The distribution of the forecasted network statistics (box plots) and the observed network statistics values (red lines) for increment (Inc) and decrement (Dec) processes from day $24$ to $28$.}
\label{fig:baboon}
\end{figure}

We choose four network statistics for this task. The estimated parameters and standard errors for both increment and decrement processes are reported in Table~\ref{tbl:Baboons}. To forecast out-of-sample data, we generate $100$ valued network $\hat{\bm{y}}^{t}$ conditional on the observed $\bm{y}^{t-1}$ for $t = 24, \cdots, 28$. Each sampled network is generated after $200 \times n \times n$ MCMC transitions starting from an empty network. We then construct the corresponding $\hat{\bm{y}}^{+,t}$ and $\hat{\bm{y}}^{-,t}$ and calculate their network statistics. The distributions of the forecasted network statistics and the observed network statistics values are displayed in Figure~\ref{fig:baboon}.

\begin{table}[!htbp]
\centering
\begin{threeparttable}
    
\begin{tabular}{ l  r  r }
\hline
\multicolumn{1}{l}{Network Statistics} & \multicolumn{1}{c}{$\bm{\eta}^+$} & \multicolumn{1}{c}{$\bm{\eta}^-$}\\
\hline
Edge sum          & \textbf{4.674} (0.017) & \textbf{-0.160} (0.014)\\
Propensity        & \textbf{9.937} (0.204) & \textbf{10.345} (0.154)\\
Dispersion        & \textbf{-14.728} (0.139) & \textbf{-14.064} (0.103)\\
Transitive weight & \textbf{-0.060} (0.007) & \textbf{-0.145} (0.007)\\
\hline
\end{tabular}

\begin{tablenotes}
\small
\item Coefficients statistically significant at $0.05$ level are bolded.
\end{tablenotes}

\end{threeparttable}
\caption{The parameter estimation (standard error) for the baboons interaction networks from day 1 to 23.}
\label{tbl:Baboons}
\end{table}

In exchange for the extrapolation of future temporal trends, the time-homogeneous PST ERGM that consolidates the fluctuation throughout $23$ days into one parameter $\bm{\eta}$ may introduce variation to the forecasted network statistics. The discrepancy on day $25$ in Figure~\ref{fig:baboon} is potentially impacted by this outcome. Furthermore, the discrepancy of the propensity term on day $27$ in the decrement process may be influenced by the increase of all network statistics from day $26$, as our proposed PST ERGM allows interaction between the two processes. Note that the $\bm{y}^{+,t}$ and $\bm{y}^{-,t}$ in PST ERGM are no longer conditionally independent as the sample space of valued networks is infinite. In summary, besides prediction error for the unseen data, the learned time-homogeneous PST ERGM effectively recovers the sudden change on day $26$ along with the temporal trends from day $24$ to $28$.

Another aspect worth mentioning is the comparison between the edge sum term and the propensity term in this experiment. The propensity term $\sum_{i<j} \mathbbm{1}(\bm{y}_{ij}>0)$ is essentially a thresholding version of the edge sum term $\sum_{i<j} \bm{y}_{ij}$. We can dichotomize a valued network into a binary network and count the number of edges to calculate the propensity term. In the first column of Figure~\ref{fig:baboon}, the observed edge sums in red lines present a decreasing trend followed by an increasing trend in both processes from day $24$ to $28$. The dispersion term and the transitivity term that are also evaluated with the valued networks produce similar patterns. However, in the second column of Figure~\ref{fig:baboon}, the observed propensity terms in red lines primarily show a decreasing trend in both processes. Empirically, dichotomizing dynamic valued networks into dynamic binary networks, or dyad value thresholding, for network analysis may introduce biases \citep{thomas2011valued} that result in unrealistic network dynamics.

\section{Discussion}
\label{sec:discussion}

This paper introduces a probabilistic model for dynamic valued networks. In practice, the factors and processes that increase relational strength are usually different from those that decrease relational strength. While dynamic network models should capture the intrinsic difference between consecutive networks, models neglecting the confounding effect of structural change may result in misinterpretation of network evolution. Inspired by \cite{STERGM}, we propose a PST ERGM to dissect valued network transitions with two sets of intermediate networks, where one manages dyad value increment, and the other manages dyad value decrement. Our proposed PST ERGM provides the interpretability of network evolution and the capability to forecast temporal trends.

Several improvements to the PST ERGM are possible for future development. We can extend the sample space to networks with continuous dyad values. In this context, novel reference functions and network statistics are needed as PST ERGM becomes a continuous probability distribution. Furthermore, besides dyad value increment and decrement, alternative ways to dissect network evolution are permitted, as long as the confounding effect of network dynamics is avoided. Over time, the number of participants and the process that induces their relations may not be fixed or completely observed. It is of great importance for a dynamic network model to identify the temporal changes punctually \citep[e.g.,][]{padilla2019change, yu2021optimal, kei2023change} and to adjust the structural changes accordingly \citep[e.g.,][]{krivitsky2011adjusting}.

Finally, model degeneracy which is studied theoretically by \cite{handcock2003assessing} is a well-known challenge in the ERGM framework. In modeling dynamic valued networks, though an infinite sample space does not have a maximal graph on which a PST ERGM will concentrate, the MLE can be difficult to find by the MCMC methods. Also, the geometrically weighted statistics that are used to alleviate the degeneracy problem in fitting static binary networks are currently not available for valued networks. Therefore, a rigorous way to design more informative network statistics as in \cite{snijders2006new}, and a systematic way to evaluate the goodness of model fit as in \cite{hunter2008goodness} are needed for dynamic valued networks. The tapered ERGM introduced in \cite{fellows2017removing}, and \cite{blackburn2022practical} can also be extended to the PST ERGM to alleviate the degeneracy issue.

\section*{Acknowledgements}

We thank the associate editor and referees for careful review and insightful comments. We also thank Mark Handcock for helpful comments on this work.

\section*{Funding}

This research did not receive any specific grant from funding agencies in the public, commercial, or not-for-profit sectors.

\section*{Data and Code}

The real data analyzed in Section \ref{subsec:friendship} is publicly available at \url{http://www.sociopatterns.org/datasets/high-school-contact-and-friendship-networks/}. The real data analyzed in Section \ref{subsec:baboons} is publicly available at \url{https://osf.io/ufs3y/}. The code to implement our methodology is publicly available at \url{https://github.com/allenkei/PSTERGM}.

\appendix

\section{Comparison of Simple Fitted Models}
\label{appendix:comparison}

In this section, we present the simple fitted models that motivate the increment and decrement networks in Section \ref{subsec:inc_dec}. First, we fit a Poisson-reference valued ERGM \citep{ValuedERGM} that involves only the edge sum term $\bm{g}(\bm{y}^t) = \sum_{(i,j)\in\mathbb{Y}} \bm{y}^t_{ij}$ to $\bm{y}^{25}$ and $\bm{y}^{26}$, respectively. The coefficients and standard errors are displayed in Table \ref{tbl:motivating}. We notice the two coefficients are similar, providing little information about the transition. Furthermore, we fit a PST ERGM that involves only the edge sum terms $\bm{g}(\bm{y}^t, \bm{y}^{t-1}) = \big(\sum_{(i,j)\in\mathbb{Y}} \bm{y}_{ij}^{+,t}, \sum_{(i,j)\in\mathbb{Y}} \bm{y}_{ij}^{-,t}\big)$ to both $\bm{y}^{+,26}$ and $\bm{y}^{-,26}$. The coefficients and standard errors are displayed in Table \ref{tbl:motivating_PSTERGM}. The positive and negative coefficients reveal the fluctuation in dyad values between the observed time points.

\begin{table}[!htbp]
\centering
\begin{threeparttable}
    
\begin{tabular}{ l  r  r }
\hline
\multicolumn{1}{l}{Network Statistics} & \multicolumn{1}{c}{$\bm{\eta}$ for $\bm{g}(\bm{y}^{25})$} & \multicolumn{1}{c}{$\bm{\eta}$ for $\bm{g}(\bm{y}^{26})$}\\
\hline
Edge sum          & \textbf{2.371} (0.033) & \textbf{2.421} (0.035)\\
\hline
\end{tabular}

\end{threeparttable}
\caption{The parameter estimation (standard error) for the baboons interaction networks on day 25 and day 26, respectively. Coefficients statistically significant at $0.05$ level are bolded.}
\label{tbl:motivating}
\end{table}

\begin{table}[!htbp]
\centering
\begin{threeparttable}
    
\begin{tabular}{ l  r  r }
\hline
\multicolumn{1}{l}{Network Statistics} & \multicolumn{1}{c}{$\bm{\eta}^+$} & \multicolumn{1}{c}{$\bm{\eta}^-$}\\
\hline
Edge sum          & \textbf{1.887} (0.055) & \textbf{-0.843} (0.069)\\
\hline
\end{tabular}

\end{threeparttable}
\caption{The parameter estimation (standard error) for the baboons interaction networks, using the increment and decrement networks. Coefficients statistically significant at $0.05$ level are bolded.}
\label{tbl:motivating_PSTERGM}
\end{table}

\section{Special Cases of PST ERGM}
\label{appendix:special_cases}

In this section, we derive two special cases of PST ERGM, under a specific set of sufficient statistics and temporal information. Consider a simple PST ERGM with the edge sums of increment and decrement networks as two network statistics:
$$\bm{g}(\bm{y}^t, \bm{y}^{t-1}) = \big(\bm{g}^+(\bm{y}^{+,t},\bm{y}^{t-1}),\bm{g}^-(\bm{y}^{-,t},\bm{y}^{t-1})\big) = \Big(\sum_{(i,j)\in\mathbb{Y}} \bm{y}_{ij}^{+,t}, \sum_{(i,j)\in\mathbb{Y}} \bm{y}_{ij}^{-,t}\Big) \in \mathbb{R}^2.$$
Let $\mathcal{Y}^t(\bm{y}^{t-1}) \subseteq \{\bm{y}^t \in \mathbb{N}_0^{\mathbb{Y}}: \bm{y}_{ij}^{t} > \bm{y}_{ij}^{t-1}\ \forall\ (i,j) \in \mathbb{Y}\}$ be a sample space for $\bm{y}^t$ starting from $\bm{y}^{t-1}$. The increment network $\bm{y}^{+,t}$ is essentially the $\bm{y}^t$, and we have
\begin{align*}
P(\bm{y}^{t}|\bm{y}^{t-1};\bm{\eta}) & = \frac{ \big[ \prod_{ij} (\bm{y}_{ij}^{t}!)^{-1} \exp(\bm{\eta}^{+} \cdot \bm{y}_{ij}^{t}) \big] \times \big[ \prod_{ij} \binom{m^t}{\bm{y}^{t-1}_{ij}} \exp(\bm{\eta}^{-} \cdot \bm{y}_{ij}^{t-1})  \big]}{  \sum_{\bm{y}^t \in \mathcal{Y}^t(\bm{y}^{t-1})}  \big[\prod_{ij} (\bm{y}_{ij}^{t}!)^{-1} \exp(\bm{\eta}^{+} \cdot \bm{y}_{ij}^{t}) \times   \prod_{ij} \binom{m^t}{\bm{y}^{t-1}_{ij}} \exp(\bm{\eta}^{-} \cdot \bm{y}_{ij}^{t-1}) \big]}\\ 
& = \frac{  \prod_{ij} \binom{m^t}{\bm{y}^{t-1}_{ij}} \exp(\bm{\eta}^{-} \cdot \bm{y}_{ij}^{t-1})}{\prod_{ij} \binom{m^t}{\bm{y}^{t-1}_{ij}} \exp(\bm{\eta}^{-} \cdot \bm{y}_{ij}^{t-1})} \times
\frac{\prod_{ij} (\bm{y}_{ij}^{t}!)^{-1} \exp(\bm{\eta}^{+} \cdot \bm{y}_{ij}^{t})}{   \sum_{\bm{y}^t \in \mathcal{Y}^t(\bm{y}^{t-1})}  \big[\prod_{ij} (\bm{y}_{ij}^{t}!)^{-1} \exp(\bm{\eta}^{+} \cdot \bm{y}_{ij}^{t}) \big]}\\
& = \prod_{ij} \frac{(\bm{y}_{ij}^{t}!)^{-1} \exp(\bm{\eta}^{+} \cdot \bm{y}_{ij}^{t})}{
\sum_{u = \bm{y}_{ij}^{t-1}+1}^{\infty}  \big[(u)^{-1} \exp(\bm{\eta}^{+} \cdot u) \big]}\\
& =\prod_{ij} \frac{ (\bm{y}_{ij}^{t}!)^{-1} \exp(\bm{\eta}^{+} \cdot \bm{y}_{ij}^{t})}{\exp\big(\exp(\bm{\eta}^{+})\big) -  \sum_{u=0}^{\bm{y}_{ij}^{t-1}} (u!)^{-1} \exp(\bm{\eta}^{+} \cdot u)}\\
&=\prod_{ij}\frac{P_{\text{Pois}}\big(\bm{y}_{ij}^t\big)}{1 -  \sum_{u=0}^{\bm{y}_{ij}^{t-1}} P_{\text{Pois}}(u)},
\end{align*}
where $P_{\text{Pois}}(x)$ denotes the probability mass function of ${\text{Poisson}}\big(\lambda = \exp(\bm{\eta}^+)\big)$ evaluated at $x$. The PST ERGM $P(\bm{y}^{t}|\bm{y}^{t-1};\bm{\eta})$ for $\bm{y}^t \in \mathcal{Y}^t(\bm{y}^{t-1})$ becomes a dyadic independent truncated Poisson distribution.

Moreover, let $\mathcal{Y}^t(\bm{y}^{t-1}) \subseteq \{\bm{y}^t \in \mathbb{N}_0^{\mathbb{Y}}: \bm{y}_{ij}^{t} < \bm{y}_{ij}^{t-1} \leq m^t \ \forall\ (i,j) \in \mathbb{Y}\}$ be another sample space for $\bm{y}^t$. The decrement network $\bm{y}^{-,t}$ is essentially the $\bm{y}^t$, and we have
\begin{align*}
P(\bm{y}^{t}|\bm{y}^{t-1};\bm{\eta}) & = \frac{ \big[ \prod_{ij} (\bm{y}_{ij}^{t-1}!)^{-1} \exp(\bm{\eta}^{+} \cdot \bm{y}_{ij}^{t-1}) \big] \times \big[\prod_{ij} \binom{m^t}{\bm{y}^{t}_{ij}} \exp(\bm{\eta}^{-} \cdot \bm{y}_{ij}^{t})  \big]}{ \sum_{\bm{y}^t \in \mathcal{Y}^t(\bm{y}^{t-1})} \big[  \prod_{ij} (\bm{y}_{ij}^{t-1}!)^{-1} \exp(\bm{\eta}^{+} \cdot \bm{y}_{ij}^{t-1}) \times \prod_{ij} \binom{m^t}{\bm{y}^{t}_{ij}} \exp(\bm{\eta}^{-} \cdot \bm{y}_{ij}^{t})\big]}\\ 
& = \frac{\prod_{ij} (\bm{y}_{ij}^{t-1}!)^{-1} \exp(\bm{\eta}^{+} \cdot \bm{y}_{ij}^{t-1})}{\prod_{ij} (\bm{y}_{ij}^{t-1}!)^{-1} \exp(\bm{\eta}^{+} \cdot \bm{y}_{ij}^{t-1})} \times
\frac{\prod_{ij} \binom{m^t}{\bm{y}^{t}_{ij}} \exp(\bm{\eta}^{-} \cdot \bm{y}_{ij}^{t})}{ \sum_{\bm{y}^t \in \mathcal{Y}^t(\bm{y}^{t-1})} \big[\prod_{ij} \binom{m^t}{\bm{y}^{t}_{ij}} \exp(\bm{\eta}^{-} \cdot \bm{y}_{ij}^{t}) \big]}\\
& = \prod_{ij} \frac{\binom{m^t}{\bm{y}^{t}_{ij}} \exp(\bm{\eta}^{-} \cdot \bm{y}_{ij}^{t})}{
 \sum_{u = 0}^{\bm{y}_{ij}^{t-1}-1} \big[\binom{m^t}{u} \exp(\bm{\eta}^{-} \cdot u) \big]}\\
& =\prod_{ij} \frac{ \binom{m^t}{\bm{y}^{t}_{ij}} \exp(\bm{\eta}^{-} \cdot \bm{y}_{ij}^{t})}{ \big(1+\exp(\bm{\eta}^-)\big)^{m^t} - \sum_{u=\bm{y}_{ij}^{t-1}}^{m^t}  \binom{m^t}{u}\exp(\bm{\eta}^{-} \cdot u)}\\
&=\prod_{ij} \frac{P_{\text{Bino}}\big(\bm{y}_{ij}^t\big)}{1 -  \sum_{u=\bm{y}_{ij}^{t-1}}^{m^t} P_{\text{Bino}}\big(u\big)},
\end{align*}
where $P_{\text{Bino}}(x)$ denotes the probability mass function of ${\text{Binomial}}\big(m^t, p = \text{logit}^{-1}(\bm{\eta}^-)\big)$ evaluated at $x$. The PST ERGM $P(\bm{y}^{t}|\bm{y}^{t-1};\bm{\eta})$ for $\bm{y}^t \in \mathcal{Y}^t(\bm{y}^{t-1})$ becomes a dyadic independent truncated Binomial distribution.

\section{Parameter Estimation Algorithms}
\label{appendix:algo}

The partial stepping algorithm of PST ERGM parameter estimation to seed an initial configuration is provided in Algorithm \ref{alg:partial_stepping_algorithm}. In this work, we let $\gamma_c^t$ be the ratio of the current iteration $c$ to the maximum number of iterations $C$ for each time $t$. Only in the last learning iteration where $\gamma_c^t = 1$ do we use the difference between observed network statistics $\sum_{t=2}^T \bm{g}(\bm{y}^{t},\bm{y}^{t-1})$ and estimated network statistics $\sum_{t=2}^T \bm{\mu}_{t}$ to update the parameter.

\begin{algorithm}
\caption{Partial stepping algorithm}
\label{alg:partial_stepping_algorithm}
\begin{algorithmic}[1]

\STATE{\textbf{Input}: initialized parameter $\bm{\eta}_0$, learning iteration $C$, sample size $s$, \{$\bm{y}^{1}, \cdots, \bm{y}^{T}$\}}

\FOR{$c = 1,\cdots, C$}

\FOR{$t = 2, \cdots, T$}

\STATE{Generate $s$ MCMC samples $\bm{y}^{t}_{1}, \cdots, \bm{y}^{t}_{s}$ from $P(\bm{y}^{t}|\bm{y}^{t-1};\bm{\eta}_{c-1})$ as in Section \ref{subsec:MH}}

\STATE{Calculate $\bm{\mu}_t = s^{-1} \sum_{i'=1}^s \bm{g}(\bm{y}_{i'}^{t},\bm{y}^{t-1})$}

\STATE{Calculate $\bm{\Sigma}_t = s^{-1} \sum_{i'=1}^s \bm{g}(\bm{y}_{i'}^{t},\bm{y}^{t-1})\bm{g}(\bm{y}_{i'}^{t},\bm{y}^{t-1})^{\top} - \bm{\mu}_t \bm{\mu}_t^{\top}$}

\ENDFOR

\STATE{For $\gamma_c^t = c/C$, calculate $\hat{\bm{\xi}}(\bm{y}) = \sum_{t=2}^T \gamma_c^t\  \bm{g}(\bm{y}^{t},\bm{y}^{t-1}) + \sum_{t=2}^T (1 - \gamma_c^t)\bm{\mu}_{t}$}

\STATE{$\bm{\eta}_{c} = \bm{\eta}_{c-1} + [\sum_{t=2}^T \bm{\Sigma}_{t}]^{-1} [\hat{\bm{\xi}}(\bm{y}) - \sum_{t=2}^T \bm{\mu}_{t}]$}

\ENDFOR

\STATE{$\tilde{\bm{\eta}} \leftarrow \bm{\eta}_{c}$}

\STATE{\textbf{Output}: learned parameter $\tilde{\bm{\eta}}$}
\end{algorithmic}
\end{algorithm}

Once a parameter is obtained from maximizing the approximated log-likelihood ratio, we proceed to the Newton-Raphson method to further update the parameter near its convergence. The Newton-Raphson algorithm of PST ERGM parameter estimation is provided in Algorithm \ref{alg:Newton_Raphson}. The $\text{CD}_K$ sampling algorithm to generate a single sampled network $\bm{y}_{i'}^{t}$ is provided in Algorithm \ref{alg:CD}, and it is used in Step $4$ of Algorithm~\ref{alg:partial_stepping_algorithm} and Step $4$ of Algorithm~\ref{alg:Newton_Raphson}.

\begin{algorithm}
\caption{Newton-Raphson algorithm}
\label{alg:Newton_Raphson}
\begin{algorithmic}[1]

\STATE{\textbf{Input}: initialized parameter $\bm{\eta}_0$, learning iteration $C$, sample size $s$, \{$\bm{y}^{1}, \cdots, \bm{y}^{T}$\}}

\FOR{$c = 1,\cdots, C$}

\FOR{$t = 2, \cdots, T$}

\STATE{Generate $s$ MCMC samples $\bm{y}^{t}_{1}, \cdots, \bm{y}^{t}_{s}$ from $P(\bm{y}^{t}|\bm{y}^{t-1};\bm{\eta}_{c-1})$ as in Section \ref{subsec:MH}}

\STATE{Calculate $\bm{\mu}_t = s^{-1} \sum_{i'=1}^s \bm{g}(\bm{y}_{i'}^{t},\bm{y}^{t-1})$}

\STATE{Calculate $\bm{\Sigma}_t = s^{-1} \sum_{i'=1}^s \bm{g}(\bm{y}_{i'}^{t},\bm{y}^{t-1})\bm{g}(\bm{y}_{i'}^{t},\bm{y}^{t-1})^{\top} - \bm{\mu}_t \bm{\mu}_t^{\top}$}

\ENDFOR

\STATE{Calculate $\hat{\bm{S}}(\bm{\eta}_{c-1}) = \sum_{t=2}^T [\bm{g}(\bm{y}^{t},\bm{y}^{t-1}) - \bm{\mu}_t]$ and $\hat{\bm{H}}(\bm{\eta}_{c-1}) = -\sum_{t=2}^T \bm{\Sigma}_t$}

\STATE{$\bm{\eta}_{c} = \bm{\eta}_{c-1} - \hat{\bm{H}}(\bm{\eta}_{c-1})^{-1} \hat{\bm{S}}(\bm{\eta}_{c-1})$}

\ENDFOR

\STATE{$\tilde{\bm{\eta}} \leftarrow \bm{\eta}_{c}$}

\STATE{\textbf{Output}: learned parameter $\tilde{\bm{\eta}}$}
\end{algorithmic}
\end{algorithm}

\begin{algorithm}
\caption{Contrastive Divergence sampling}
\label{alg:CD}
\begin{algorithmic}[1]

\STATE{\textbf{Input}: MCMC transition step $K$, parameter $\bm{\eta} = (\bm{\eta}^+,\bm{\eta}^-)$, \{$\bm{y}^{t-1}, \bm{y}^{t}$\}}

\STATE{Set $\tilde{\bm{y}} = \bm{y}^t$}

\FOR{$k = 1,\cdots, K$}

\STATE{Choose randomly a dyad $(i,j)$ s.t. $i \neq j$}

\STATE{Propose a dyad value $\tilde{\bm{y}}^{k+1}_{ij}$ from Equation (\ref{eq:proposed})}

\STATE{Calculate the acceptance ratio $\alpha$ for $\tilde{\bm{y}}^{k+1}_{ij}$ from Equation (\ref{eq:acceptance})}

\IF{$\text{uniform}(0,1) < \alpha$}

\STATE{Accept $\tilde{\bm{y}}^{k+1}_{ij}$}

\ENDIF

\ENDFOR

\STATE{\textbf{Output}: a sampled network $\tilde{\bm{y}}$}
\end{algorithmic}
\end{algorithm}

The complexity of Algorithm~\ref{alg:CD} to sample one valued network with $K$ MCMC transitions is $O(K\mathcal{C}_{\alpha})$. The term $\mathcal{C}_{\alpha}$ is the complexity of calculating the acceptance ratio $\alpha$ with Equation~(\ref{eq:acceptance}), which depends on the choice of network statistics for both increment and decrement processes as well as the randomness of the proposed dyad values from Equation~(\ref{eq:proposed}). Likewise, for Algorithm~\ref{alg:partial_stepping_algorithm}, the complexity to sample $s$ valued networks in Step 4 is $O(sK\mathcal{C}_{\alpha})$. In Steps $5$ and $6$ of Algorithm~\ref{alg:partial_stepping_algorithm}, the complexity of calculating the mean vector $\bm{\mu}_t \in \mathbb{R}^p$ and the covariance matrix $\bm{\Sigma}_t \in \mathbb{R}^{p \times p}$ of $s$ sampled network statistics is $O(s\mathcal{C}_{2g} + sp + sp^2)$. The term $\mathcal{C}_{2g}$ is the complexity to calculate the two network statistics $\bm{g}(\bm{y}^{t},\bm{y}^{t-1})$ that are based on the user's choice. Finally, the complexity of calculating the pseudo-observation $\hat{\bm{\xi}}(\bm{y})$ and the maximizer $\tilde{\bm{\eta}}$ in Steps $8$ and $9$ of Algorithm~\ref{alg:partial_stepping_algorithm} is $O(T\mathcal{C}_{2g} + T + p^2)$, where $T$ is the number of observed networks. Overall, the complexity of Algorithm~\ref{alg:partial_stepping_algorithm} is $O\big(C[T(sK\mathcal{C}_{\alpha}+s\mathcal{C}_{2g} + sp + sp^2) + T\mathcal{C}_{2g} + T + p^2]\big)$, where $C$ is the number of learning iterations. The complexity of Algorithm~\ref{alg:Newton_Raphson} is similar, except for different choices of the input parameters and MCMC variation in the proposed dyad values.

\section{Experiment Details}
\label{appendix:exp}

In this section, we provide the formulations of selected network statistics used in the two real data examples. The network statistics of interest are chosen from an extensive list in \texttt{ergm} \citep{ergm_package}, an \texttt{R} library for network analysis. The choice of network statistics for the increment process is identical to that for the decrement process. Furthermore, we provide the learning schedules of the two experiments.

\subsection{Students Contact Networks}

The formulations of six network statistics used for the students contact networks \citep{mastrandrea2015contact} are displayed in Table~\ref{tbl:Friendship_data_network_statistics}. The superscripts are omitted for notational simplicity. In this time-heterogeneous PST ERGM $\prod_{t=2}^5 P(\bm{y}^t|\bm{y}^{t-1};\bm{\eta}^t)$, the parameter $\bm{\eta}^t$ is learned sequentially for $t = 2, \cdots, 5$, and  we initialize $\bm{\eta}_0^t$ of Algorithm~\ref{alg:partial_stepping_algorithm} as zero vector. 

\begin{table}[!htbp]
\centering
\begin{threeparttable}
    
\begin{tabular}{ l  l }
\hline
\multicolumn{1}{l}{Network Statistics} & \multicolumn{1}{l}{Formulations}\\
\hline
Edge sum & $\sum_{i<j} \bm{y}_{ij}$ \\
Dispersion & $\sum_{i<j} \sqrt{\bm{y}_{ij}}$\\
Homophily (M) & $\sum_{i<j} \bm{y}_{ij} \times \mathbbm{1}(\bm{x}_i = \text{M} \land \bm{x}_j = \text{M})$ \\
Heterophily (M-F) & $\sum_{i<j} \bm{y}_{ij} \times \mathbbm{1}(\bm{x}_i \neq \bm{x}_j)$ \\
Facebook & $\sum_{i<j} \bm{y}_{ij} \times \bm{z}_{ij}$ \\
Transitive weight & $\sum_{i<j} \min (\bm{y}_{ij}, \max_{k \in N}(\min(\bm{y}_{ik},\bm{y}_{kj})))$\\
\hline
\end{tabular}

\end{threeparttable}
\caption{The network statistics used for the students contact networks.}
\label{tbl:Friendship_data_network_statistics}
\end{table}

To seed an initial configuration for each $\bm{\eta}^t$, we implement $C = 20$ iterations of Algorithm \ref{alg:partial_stepping_algorithm} where an MCMC sample size of $s = 100$ with $\text{CD}_{5 \times n}$ sampling is used, followed by another $C = 20$ iterations of Algorithm \ref{alg:Newton_Raphson} where an MCMC sample size of $s = 100$ with $\text{CD}_{10 \times n}$ sampling is used. The term $n$ is the number of students in this data, which is $29$. Subsequently, to refine the learned parameters, we implement $C = 10$ iterations of Algorithm \ref{alg:Newton_Raphson} where an MCMC sample size of $s = 1000$ with $\text{CD}_{25 \times n \times n}$ sampling is used. 

Finally, the standard errors are obtained from the Fisher Information matrix $\bm{I}(\bm{\eta}^t) \approx - \hat{\bm{H}}(\tilde{\bm{\eta}}^t)$ of Equation (\ref{eq:Hessian}) evaluated at the learned parameter $\tilde{\bm{\eta}}^t$ with $1000$ sampled networks. Each sampled network is generated after $K = 20 \times n \times n$ MCMC transitions starting from observed $\bm{y}^t$.

\subsection{Baboons Interaction Networks}

The formulations of four network statistics used for the baboons interaction networks \citep{gelardi2020measuring} are displayed in Table~\ref{tbl:Baboons_data_network_statistics}. The superscripts are omitted for notational simplicity. In this time-homogeneous PST ERGM $\prod_{t=2}^{23} P(\bm{y}^t|\bm{y}^{t-1})$, the parameter $\bm{\eta}$ is shared across $t = 2, \cdots, 23$ and is also used to forecast the temporal trends for $t = 24, \cdots, 28$. We initialize $\bm{\eta}_0$ of Algorithm~\ref{alg:partial_stepping_algorithm} as zero vector.

\begin{table}[!htbp]
\centering
\begin{threeparttable}
    
\begin{tabular}{ l  l }
\hline
\multicolumn{1}{l}{Network Statistics} & \multicolumn{1}{l}{Formulations}\\
\hline
Edge sum & $\sum_{i<j} \bm{y}_{ij}$ \\
Propensity & $\sum_{i<j} \mathbbm{1}(\bm{y}_{ij}>0)$ \\
Dispersion & $\sum_{i<j} \sqrt{\bm{y}_{ij}}$ \\
Transitive weight & $\sum_{i<j} \min (\bm{y}_{ij}, \max_{k \in N}(\min(\bm{y}_{ik},\bm{y}_{kj})))$ \\
\hline
\end{tabular}

\end{threeparttable}
\caption{The network statistics used for the baboons interaction networks.}
\label{tbl:Baboons_data_network_statistics}
\end{table}

To seed an initial configuration for $\bm{\eta}$, we implement $C = 20$ iterations of Algorithm \ref{alg:partial_stepping_algorithm} where an MCMC sample size of $s = 100$ with $\text{CD}_n$ sampling is used for each time $t$, followed by another $C = 20$ iterations of Algorithm \ref{alg:Newton_Raphson} where an MCMC sample size of $s = 100$ with $\text{CD}_{2 \times n}$ sampling is used for each time $t$. The term $n$ is the number of baboons in this data, which is $13$. Subsequently, to refine the learned parameter, we implement $C = 10$ iterations of Algorithm \ref{alg:Newton_Raphson} where an MCMC sample size of $s = 1000$ with $\text{CD}_{50 \times n \times n}$ sampling is used for each time $t$. In this experiment, we let the maximum dyad value of decrement networks $m^t=200$ for each $t$, a moderate upper bound that is greater than the highest dyad value of decrement networks $\bm{y}^{-,2}, \cdots, \bm{y}^{-,28}$ constructed from the observed networks.

Finally, the standard errors are obtained from the Fisher Information matrix $\bm{I}(\bm{\eta}) \approx - \hat{\bm{H}}(\tilde{\bm{\eta}})$ of Equation (\ref{eq:Hessian}) evaluated at the learned parameter $\tilde{\bm{\eta}}$ with $1000$ sampled networks for each time $t$. Each sampled network is generated after $K = 20 \times n \times n$ MCMC transitions starting from observed $\bm{y}^t$.


\newpage 
\bibliographystyle{elsarticle-harv} 
\bibliography{references}

\end{document}